\documentclass[twocolumn,seceq]{jpsj2}

\newcommand{\E}{\mathrm{e}}
\newcommand{\I}{\mathrm{i}}
\newcommand{\D}{\mathrm{d}}

\title{Strong Resonance of Light in a Cantor Set}

\author{Naomichi \textsc{Hatano}\thanks{E-mail address: hatano@iis.u-tokyo.ac.jp}}
\inst{Institute of Industrial Science, University of Tokyo, Komaba, Meguro, Tokyo 153-8505}

\abst{The propagation of an electromagnetic wave in a one-dimensional fractal object, the Cantor set, is studied.
The transfer matrix of the wave amplitude is formulated and its renormalization transformation is analyzed.
The focus is on resonant states in the Cantor set.
In Cantor sets of higher generations, some of the resonant states closely approach the real axis of the wave number, leaving between them a wide region free of resonant states.
As a result, wide regions of nearly total reflection appear with sharp peaks of the transmission coefficient beside them.
It is also revealed that the electromagnetic wave is strongly enhanced and localized in the cavity of the Cantor set near the resonant frequency.
The enhancement factor of the wave amplitude at the resonant frequency is approximately $6/\left|\eta_\mathrm{r}\right|$, where $\eta_\mathrm{r}$ is the imaginary part of the corresponding resonant eigenvalue.
For example, a resonant state of the lifetime $\tau_\mathrm{r}=4.3$ms and of the enhancement factor $M=7.8\times10^7$ is found at the resonant frequency $\omega_\mathrm{r}=367$GHz for the Cantor set of the fourth generation of length $L=10$cm made of a medium of the dielectric constant $\varepsilon=10$.}

\kword{microwave, millimetric wave, terahertz light, resonance, localization, fractal, Cantor set}

\begin{document}
\maketitle

\section{Introduction}
\label{sec-intro}
We study in the present paper the propagation of an electromagnetic wave in a fractal object.
Takeda \textit{et al}.\ recently reported an interesting experiment\cite{Takeda} in which they put an electromagnetic microwave into Menger sponges made of a dielectric substance (epoxy resin).
The Menger sponge\cite{Mandelbrot} is a fractal object of the similarity dimension $\log 20/\log 3\simeq 2.7$, embedded in three spatial dimensions.
They set the fractal medium in a square tube, put an incident wave from one end of the tube, and measured the reflection wave on the same end as well as the transmission wave on the other end.
They observed at $\omega=12.8$GHz narrow dips both in the transmission amplitude and the reflection amplitude and claimed a strong localization of light at the frequency.
We show in the present paper that:
\begin{enumerate}
\item the strong localization of light indeed occurs in a fractal object at various frequencies;
\item resonant states with long lifetimes are responsible for the strong localization.
\end{enumerate}
We thereby suggest that, in their experiment, the microwave of frequency $12.8$GHz was diverted to all directions because of resonant scattering and hence the amplitudes at the both ends of the tube were lost.

\begin{figure}
\begin{center}
\includegraphics[width=0.8\columnwidth]{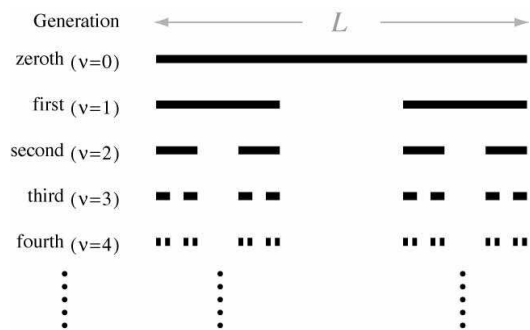}
\end{center}
\caption{The recursive construction of the Cantor set.
We first prepare a line segment of length $L$, which we refer to as the zeroth generation ($\nu=0$) of the Cantor set.
The first generation ($\nu=1$) of the Cantor set is constructed by dividing the zeroth generation into three segments and removing the central one.
The second generation ($\nu=2$) is constructed by dividing each segment of the first generation into three segments and removing the central one of each.
The Cantor sets of higher generations are thus constructed recursively.
In the strict sense of the term, the Cantor set as a fractal object is the one obtained in the limit $\nu\to\infty$;
the objects of finite $\nu$ should be called prefractals.
We here use the term rather loosely and refer to the object even of a finite $\nu$ as a Cantor set of the $\nu$th generation.}
\label{Cantor}
\end{figure}

More specifically, we theoretically calculate the wave propagation in the Cantor set.
The Cantor set\cite{Mandelbrot} (Fig.~\ref{Cantor}) is another fractal object of the similarity dimension $\log 2/\log3\simeq 0.63$, embedded in one spatial dimension.
In order to realize the Cantor set experimentally, one can use layers of sheets of dielectric media and put the light normal to the sheets.

The transmission and reflection coefficients of electromagnetic waves in the Cantor set was first computed by Sun and Jaggard in 1991 with scattering matrices,\cite{Sun91} later by Bertolotti \textit{et al.} in 1996 with transfer matrices\cite{Bertolotti} and recently by Yamanaka and Kohmoto again with transfer matrices.\cite{Yamanaka}
The novel point of the present paper is to attribute peaks and dips of the transmission and reflection coefficients to resonance poles in the complex wave-number plane;
to our knowledge, resonant states of electromagnetic waves in the Cantor set have never been analyzed.
Ohtaka\cite{Ohtaka04,Ohtaka-book} recently derived from the scattering matrix of a finite system, the density of states, which is closely related to the resonance poles.
It may be interesting to apply his formula to a Cantor set.

We first in \S\ref{sec-transfer} formulate the transfer-matrix method of computing wave amplitudes.
The transfer matrix for a Cantor set of the $\nu$th generation, $T^{(\nu)}(\zeta)$, is defined by the transfer matrix for a Cantor set of the $(\nu-1)$th generation, $T^{(\nu-1)}(\zeta)$, where $\zeta$ denotes the wave number of the incident light, made dimensionless by the system size $L$.
Thus we obtain the transfer matrix for a Cantor set of an arbitrary generation recursively.
Incidentally, it is common\cite{Pendry} in analyzing photonic crystals to use the scattering matrix instead of the transfer matrix;
the transfer matrix, when we multiply it by itself many times, can be numerically divergent.
In the present case, however, we multiply transfer matrices hierarchically and hence the numerical instability does not occur.

Once the transfer matrix is given, we compute the transmission coefficient $T$ and the reflection coefficient $R$ from matrix elements of the transfer matrix as
\begin{eqnarray}\label{eq010}
T(\zeta)&=&\left|\left(T^{(\nu)}(\zeta)\right)_{22}\right|^{-2},
\\ \label{eq020}
R(\zeta)&=&T\times\left|\left(T^{(\nu)}(\zeta)\right)_{12}\right|^{2}.
\end{eqnarray}
We also define the resonance poles $\zeta_\mathrm{r}=\xi_\mathrm{r}+\I\eta_\mathrm{r}$ as the zeros of the lower right element of the transfer matrix:
\begin{equation}\label{eq030}
\left(T^{(\nu)}(\zeta_\mathrm{r})\right)_{22}=0.
\end{equation}

In \S\ref{sec-numerical}, we demonstrate by numerical calculation that the reflection coefficient $R$ and the transmission coefficient $T$ oscillate as we vary the frequency of the incident wave (Fig.~\ref{figTandR} below).
The oscillation is very rapid particularly for media of large dielectric constants and for Cantor sets of higher generations.\cite{Sun91,Bertolotti,Yamanaka}

We also demonstrate that the resonance poles $\zeta_\mathrm{r}=\xi_\mathrm{r}+\I\eta_\mathrm{r}$ appear intermittently along the real axis of the (dimensionless) wave number $\zeta$;
these resonant states are responsible for the rapid oscillation of the reflection and transmission coefficients.
Each peak of the transmission coefficient (and hence each dip of the reflection coefficient) corresponds to a resonance pole nearby on the complex wave-number plane.

The merit of considering the resonant state is the use of the imaginary part of the resonant eigenvalue, $\eta_\mathrm{r}$.
The inverse of the imaginary part gives the resonant lifetime in the form
\begin{equation}\label{eq040}
\tau_\mathrm{r}=\frac{L/c_1}{\left|\eta_\mathrm{r}\right|},
\end{equation}
where $L$ is the entire length of the Cantor set and $c_1$ is the speed of light outside the Cantor set.
If we put an incident wave packet of nearly the resonant frequency $\omega_\mathrm{r}=\xi_\mathrm{r}\times c_1/L$, the wave packet stays inside the cavity of the Cantor set for as long as the resonant lifetime $\tau_\mathrm{r}$.

We also reveal that the inverse of the imaginary part of the resonant eigenvalue has another indication for the standing wave.
The standing wave resonates and is strongly enhanced inside the cavity of the Cantor set.
The enhancement factor $M$ (the magnitude of the standing wave inside the cavity divided by the magnitude of the incident wave) is shown to be
\begin{equation}\label{eq050}
M\simeq\frac{6}{\left|\eta_\mathrm{r}\right|}.
\end{equation}

Some of the resonant states have very long lifetimes, again particularly for media of large dielectric constants and for Cantor sets of higher generations.
For example, a Cantor set of the fourth generation $\nu=4$ with the dielectric constant $\varepsilon=10$ and of length $L=10$cm has a resonant state of lifetime $\tau_\mathrm{r}=4.3$ms at the resonant frequency $\omega_\mathrm{r}=367$GHz (Table~\ref{tableres} below).
For the wave number near such resonant states, the wave amplitude is strongly enhanced (or localized) inside the cavity of the Cantor set (Fig.~\ref{fieldres} below).
The enhancement factor $M$ in the above example is no less than $10^7$, a surprisingly large number.

\begin{figure}
\begin{center}
\includegraphics[width=0.8\columnwidth]{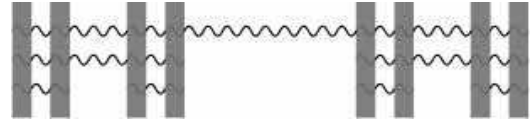}
\end{center}
\caption{A wave with the wave number $3^\nu m\pi/2L$ ($m$ is a positive integer) resonates in every cavity of the Cantor set.}
\label{cavity}
\end{figure}

This enhancement is a consequence of repeated reflection and interference inside the cavity of the Cantor set.
Roughly speaking, a wave with the wave number $3^\nu m\pi/2L$ ($m$ is a positive integer) forms a standing wave in every cavity of the Cantor set and hence has strong positive interference; see Fig.~\ref{cavity}.

In the rest of the paper, we investigate why the reflection and transmission coefficients oscillate rapidly and why some of the resonant states have very long lifetimes.
For the purpose, we introduce in \S\ref{sec-renom} a renormalization transformation of the transfer matrix.
We show that the transfer matrix for the Cantor set of any generation is expressed in the form
\begin{equation}\label{eq060}
T^{(\nu)}(\zeta)=\E^{\I\zeta(s^{(\nu)}\sigma_z-\I r^{(\nu)}\sigma_x)},
\end{equation}
where $s^{(\nu)}(\zeta)$ and $r^{(\nu)}(\zeta)$ are real functions and $\sigma_z$ and $\sigma_x$ are the Pauli matrices.
With this expression~(\ref{eq060}), we renormalize the fractal structure into a uniform medium, within which transmission $s^{(\nu)}$ and reflection $r^{(\nu)}$ occur continually.

We then analyze in \S\ref{sec-pert} the renormalization transformation
\begin{equation}\label{eq070}
\left(s^{(\nu-1)}, r^{(\nu-1)}\right)
\longrightarrow
\left(s^{(\nu)}, r^{(\nu)}\right)
\end{equation}
perturbatively for a small dielectric constant $\varepsilon\sim1$.
We observe at the dimensionless wave number $\zeta=3^\nu m\pi/2$ ($m$ is a positive integer), a rapid increase of the reflection part $r^{(\nu)}$, and consequently a rapid increase of the reflection coefficient $R$ in the form
\begin{equation}\label{eq080}
R\simeq 2^{2\nu}\lambda^2,
\end{equation}
where $\lambda=(\varepsilon-1)/2$ is the perturbation parameter.
At other wave numbers, the reflection part $r^{(\nu)}$ decreases as $\sim(2/3)^\nu$ and the system is renormalized to a trivial fixed point
\begin{equation}\label{eq090}
r^{(\nu)}\longrightarrow 0.
\end{equation}
We have the perfect transmission, $T=1$ and $R=0$, at this trivial fixed point.

\begin{figure}
\begin{center}
\includegraphics[width=0.8\columnwidth]{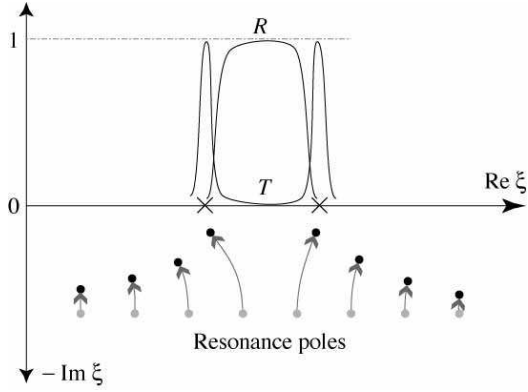}
\end{center}
\caption{For $\nu=0$, the resonance poles are regularly aligned on the lower half plane of $\zeta$.
As we progress to higher generations of the Cantor set, some of the poles move toward trivial fixed points (indicated by crosses) on the real axis, opening at the same time a wide gap between them.
}
\label{resmove}
\end{figure}

We finally analyze how the resonance poles $\zeta_\mathrm{r}$ move in the complex wave-number plane as we renormalize many times, or as we progress to higher generations of the Cantor set.
We conclude in two opposite limits (the weakly renormalized limit $\left|s^{(\nu)}\right|\gg\left|r^{(\nu)}\right|$ in \S\ref{sec-pert} and the strongly renormalized limit  $\left|s^{(\nu)}\right|\ll\left|r^{(\nu)}\right|$ in \S\ref{sec-opposite}) that, as the renormalization progresses, some of the resonant states approach trivial fixed points~(\ref{eq090}) (examined in \S\ref{sec-trivial}) on the real axis of the wave number, leaving between them a wide region free of resonant states (Fig.~\ref{resmove}).
As a result, a wide region of nearly total reflection ($R\sim1$) appears with sharp peaks of the transmission coefficient ($T\sim1$) beside them.
This is the main point of the present paper;
the fractal structure of the medium is responsible for the behavior shown in Fig.~\ref{resmove}.

Incidentally, we heavily use M.~Suzuki's quantum analysis in the perturbation analyses in \S\S\ref{sec-pert}--\ref{sec-opposite}.
We give details of the calculation in \ref{app-QA}.

\section{Transfer matrix}
\label{sec-transfer}
\setcounter{equation}{0}

In this section, we formulate the transfer matrix of wave amplitudes in Cantor sets.
We first construct the transfer matrix for a dielectric layer, and then for Cantor sets of higher generations recursively.

\begin{figure}
\hspace{0.05\columnwidth}
\begin{minipage}[t]{0.288\columnwidth}
\vspace{0pt}
\includegraphics[width=\textwidth]{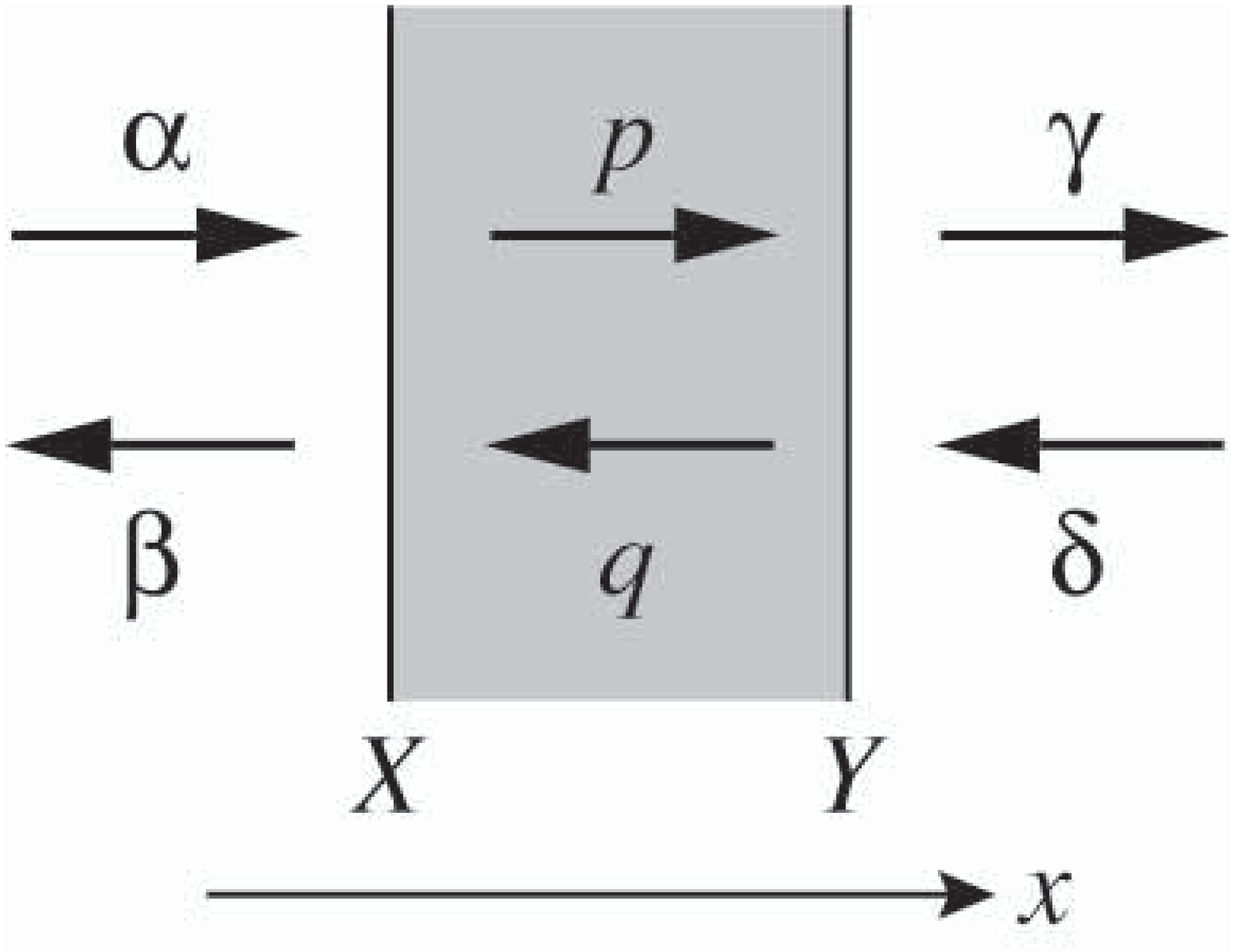}
\begin{center}(a)\end{center}
\end{minipage}
\hfill
\begin{minipage}[t]{0.508\columnwidth}
\vspace{0pt}
\includegraphics[width=\textwidth]{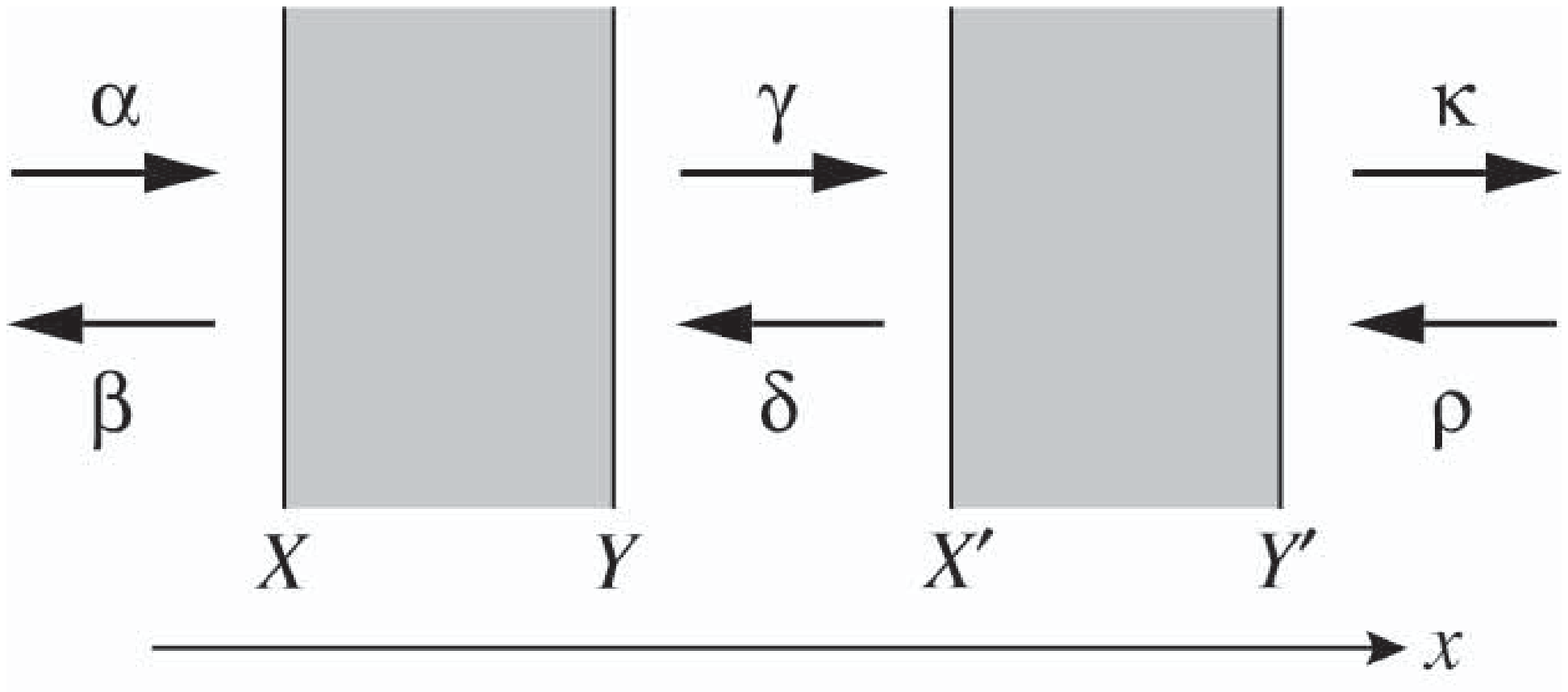}
\begin{center}(b)\end{center}
\end{minipage}
\hspace{0.05\columnwidth}
\caption{(a) A layer of the medium~2 (the gray area) sandwiched by layers of the medium~1.
(b) Two layers of the medium~2 (the gray areas) sandwiched by layers of the medium~1.}
\label{figlayers}
\end{figure}

Let us begin with solving the situation in Fig.~\ref{figlayers}~(a).
Using the Lorentz gauge, we write the Maxwell equation in the form
\begin{equation}\label{eq100}
\Delta \vec{A}=\varepsilon_i\mu_i\frac{\partial^2}{\partial t^2}\vec{A},
\end{equation}
where $\varepsilon_i$ and $\mu_i$ denote the dielectric constant and the magnetic permeability of the medium $i(=1,2)$, respectively, with the light speed
\begin{equation}\label{eq105}
c_i=\frac{1}{\sqrt{\varepsilon_i \mu_i}}.
\end{equation}
Since the two transverse components of the vector potential $\vec{A}$ are decoupled, we hereafter consider only one component $A\E^{\I\omega t}$.
Thus we have
\begin{equation}\label{eq110}
\frac{\D^2}{\D x^2}A=-\varepsilon_i\mu_i\omega^2 A.
\end{equation}
The solution of Eq.~(\ref{eq110}) is given by
\begin{equation}\label{eq120}
A(x)=\left\{
\begin{array}{ll}
\alpha\E^{\I k_1x}+\beta\E^{-\I k_1x},&
\quad\mbox{for $x<X$},\\
p\E^{\I k_2x}+q\E^{-\I k_2x},&
\quad\mbox{for $X<x<Y$},\\
\gamma\E^{\I k_1x}+\delta\E^{-\I k_1x},&
\quad\mbox{for $x>Y$},
\end{array}
\right.
\end{equation}
with $k_i=\sqrt{\varepsilon_i\mu_i}\omega=\omega/c_i$.
Note that
\begin{equation}\label{eq130}
\frac{k_2}{k_1}=\frac{\sqrt{\varepsilon_2\mu_2}}{\sqrt{\varepsilon_1\mu_1}}=\frac{c_1}{c_2}=n,
\end{equation}
where $n$ stands for the refraction constant.
The boundary conditions at $x=X$ are given by
\begin{eqnarray}\label{eq140}
A(X-0)&=&A(X+0),
\\ \label{eq141}
\frac{1}{\mu_1}A'(X-0)&=&\frac{1}{\mu_2}A'(X+0)
\end{eqnarray}
and those at $x=Y$ are given likewise.
We thus have
\begin{eqnarray}\label{eq150}
&&
\alpha\E^{\I k_1X}+\beta\E^{-\I k_1X}
=
p\E^{\I k_2X}+q\E^{-\I k_2X},\qquad
\\
\label{eq160}
&&
\alpha\frac{\I k_1}{\mu_1}\E^{\I k_1X}
-\beta\frac{\I k_1}{\mu_1}\E^{-\I k_1X}
\nonumber\\
&&
\phantom{\alpha\E^{\I k_1X}}
=
p\frac{\I k_2}{\mu_2}\E^{\I k_2X}
-q\frac{\I k_2}{\mu_2}\E^{-\I k_2X},\qquad
\end{eqnarray}
or, in a matrix form,
\begin{eqnarray}\label{eq170}
\lefteqn{
\frac{1}{\sqrt{2}}
\left(
\begin{array}{cc}
\E^{\I k_1X} & \E^{-\I k_1X} \\
\I\E^{\I k_1X} & -\I\E^{-\I k_1X}
\end{array}
\right)
\left(
\begin{array}{c}
\alpha\\
\beta
\end{array}
\right)
}
\nonumber\\
&&=
\sqrt{n'}
\left(
\begin{array}{cc}
\frac{1}{\sqrt{n'}} & \\
& \sqrt{n'}
\end{array}
\right)
\nonumber\\
&&\phantom{=}
\times
\frac{1}{\sqrt{2}}
\left(
\begin{array}{cc}
\E^{\I k_2X} & \E^{-\I k_2X} \\
\I\E^{\I k_2X} & -\I\E^{-\I k_2X}
\end{array}
\right)
\left(
\begin{array}{c}
p\\
q
\end{array}
\right),
\qquad
\end{eqnarray}
where
\begin{equation}\label{eq190}
n'=\frac{k_2}{k_1}\times\frac{\mu_1}{\mu_2}=\sqrt{\frac{\varepsilon_2\mu_1}{\varepsilon_1\mu_2}}.
\end{equation}
In reality, the magnetic permeability is almost constant for most media,
\begin{equation}\label{eq195}
\mu_1\simeq\mu_2,
\quad\mbox{and hence}\quad
n'\simeq n\simeq\sqrt{\varepsilon},
\end{equation}
where 
\begin{equation}\label{eq115}
\varepsilon\equiv\frac{\varepsilon_2}{\varepsilon_1}
\end{equation}
is the relative dielectric constant.

Applying the matrix
\begin{equation}\label{eq200}
U=\frac{1}{\sqrt{2}}
\left(
\begin{array}{cc}
1 & -\I \\
\I & -1
\end{array}
\right)
=U^{-1}
\end{equation}
from the left of the both sides of Eq.~(\ref{eq170}), we have
\begin{eqnarray}\label{eq210}
\lefteqn{
\left(
\begin{array}{cc}
\E^{\I k_1X} & \\
& \I\E^{-\I k_1X}
\end{array}
\right)
\left(
\begin{array}{c}
\alpha\\
\beta
\end{array}
\right)
}
\nonumber\\
&=&
\sqrt{n'}
\left(
\begin{array}{cc}
\cos\frac{\phi}{2} & -\sin\frac{\phi}{2} \\
\sin\frac{\phi}{2} & \cos\frac{\phi}{2}
\end{array}
\right)
\nonumber\\
&&\times
\left(
\begin{array}{cc}
\E^{\I k_2X} & \\
& \I\E^{-\I k_2X}
\end{array}
\right)
\left(
\begin{array}{c}
p\\
q
\end{array}
\right),
\end{eqnarray}
where the angle $\phi$ is defined by
\begin{equation}\label{eq230}
\E^{\I\phi}=n'.
\end{equation}
(The constant $\phi$ is in fact a purely imaginary number.)
We now express Eq.~(\ref{eq210}) in terms of the Pauli matrices as
\begin{equation}\label{eq240}
\E^{\I k_1X\sigma_z}
\left(
\begin{array}{c}
\alpha\\
\I\beta
\end{array}
\right)
=
\sqrt{n'}
\E^{-\I\phi\sigma_y/2}
\E^{\I k_2X\sigma_z}
\left(
\begin{array}{c}
p\\
\I q
\end{array}
\right).
\end{equation}
Likewise, we have
\begin{equation}\label{eq250}
\E^{\I k_1Y\sigma_z}
\left(
\begin{array}{c}
\gamma\\
\I\delta
\end{array}
\right)
=
\sqrt{n'}
\E^{-\I\phi\sigma_y/2}
\E^{\I k_2Y\sigma_z}
\left(
\begin{array}{c}
p\\
\I q
\end{array}
\right)
\end{equation}
for the boundary conditions at $x=Y$.
Solving Eq.~(\ref{eq240}) with respect to $(p,\I q)$ and substituting the solution for $(p,\I q)$ in Eq.~(\ref{eq250}), we have
\begin{eqnarray}\label{eq260}
\lefteqn{
\E^{\I k_1Y\sigma_z}
\left(
\begin{array}{c}
\gamma\\
\I\delta
\end{array}
\right)
}
\nonumber\\
&&=
\E^{-\I\phi\sigma_y/2}
\E^{\I k_2(Y-X)\sigma_z}
\E^{\I\phi\sigma_y/2}
\nonumber\\
&&
\phantom{\E^{-\I\phi\sigma_y/2}
\E^{\I k_2\sigma_z}}
\times
\E^{\I k_1X\sigma_z}
\left(
\begin{array}{c}
\alpha\\
\I\beta
\end{array}
\right)
\\
\label{eq270}
&&=
\E^{\I nk_1(Y-X)(\sigma_z\cos\phi+\sigma_x\sin\phi)}
\nonumber\\
&&
\phantom{\E^{\I nk_1(\sigma_z\cos\phi)}}
\times
\E^{\I k_1X\sigma_z}
\left(
\begin{array}{c}
\alpha\\
\I\beta
\end{array}
\right).
\end{eqnarray}
Hence we define the transfer matrix for the medium~2 as
\begin{equation}\label{eq280}
T_2(\zeta)\equiv
\E^{-\I\phi\sigma_y/2}
\E^{\I \zeta n\sigma_z}
\E^{\I\phi\sigma_y/2}
=
\E^{\I\zeta n(\sigma_z\cos\phi+\sigma_x\sin\phi)}.
\end{equation}
The coefficients of the Pauli matrices in the exponent are
\begin{eqnarray}\label{eq285}
&&n\cos\phi=
\frac{1}{2}\left(\frac{\varepsilon_2}{\varepsilon_1}+\frac{\mu_2}{\mu_1}\right)
\simeq\frac{1}{2}\left(\varepsilon+1\right),
\qquad
\\
\label{eq286}
&&n\sin\phi=
\frac{1}{2\I}\left(\frac{\varepsilon_2}{\varepsilon_1}-\frac{\mu_2}{\mu_1}\right)
\simeq\frac{1}{2\I}\left(\varepsilon-1\right).
\qquad
\end{eqnarray}

The situation in Fig.~\ref{figlayers}~(b) is then described by
\begin{eqnarray}\label{eq290}
\lefteqn{
\E^{\I k_1Y'\sigma_z}
\left(
\begin{array}{c}
\kappa\\
\I\rho
\end{array}
\right)
=
T_2(k_1(Y'-X'))
\E^{\I k_1X'\sigma_z}
\left(
\begin{array}{c}
\gamma\\
\I\delta
\end{array}
\right)
}
\nonumber\\
&&=
T_2(k_1(Y'-X'))
\E^{\I k_1(X'-Y)\sigma_z}
\E^{\I k_1Y\sigma_z}
\left(
\begin{array}{c}
\gamma\\
\I\delta
\end{array}
\right)
\nonumber\\
&&=
T_2(k_1(Y'-X'))
\E^{\I k_1(X'-Y)\sigma_z}
T_2(k_1(Y-X))
\nonumber\\
&&\qquad
\times
\E^{\I k_1X\sigma_z}
\left(
\begin{array}{c}
\alpha\\
\I\beta
\end{array}
\right).
\end{eqnarray}
Hence we define the transfer matrix for the medium 1 as
\begin{equation}\label{eq300}
T_1(\zeta)\equiv
\E^{\I\zeta\sigma_z}.
\end{equation}

We are now in the position of writing down the transfer matrix for the Cantor set.
The transfer matrix for the Cantor set of the first generation is given by
\begin{equation}\label{eq310}
T^{(1)}(\zeta)
=T_2\left(\frac{\zeta}{3}\right)
T_1\left(\frac{\zeta}{3}\right)
T_2\left(\frac{\zeta}{3}\right),
\end{equation}
where
\begin{equation}\label{eq311}
\zeta=k_1 L
\end{equation}
is a dimensionless wave number with $L$ being the length of the Cantor set.
We thereby generate the transfer matrix for Cantor sets of higher generations recursively as
\begin{equation}\label{eq320}
T^{(\nu+1)}(\zeta)
=T^{(\nu)}\left(\frac{\zeta}{3}\right)
T_1\left(\frac{\zeta}{3}\right)
T^{(\nu)}\left(\frac{\zeta}{3}\right),
\end{equation}
where $\nu$ denotes the generation of the Cantor set.
We also define
\begin{equation}\label{eq325}
T^{(0)}(\zeta)\equiv T_2(\zeta),
\end{equation}
which makes Eq.~(\ref{eq310}) a part of the recursion relation~(\ref{eq320}).

Finally, we mention two symmetries of the transfer matrix.
First, because the Pauli matrices $\sigma_z$ and $\sigma_x$ are both symmetric, the transfer matrices for the media 1 and 2, Eqs.~(\ref{eq280}) and~(\ref{eq300}), are also symmetric.
The transfer matrices for the Cantor sets of all generations are therefore symmetric as well:
\begin{equation}\label{eq330}
{}^\mathrm{t}T^{(\nu)}=T^{(\nu)}.
\end{equation}
Next, we note the fact that $\det \E^{a\sigma_\mu}=1$ ($\mu=x,y,z$) for an arbitrary coefficient $a$.
Because the transfer matrix $T^{(\nu)}$ is written as a product of exponential operators of the Pauli matrices, we conclude that
\begin{equation}\label{eq340}
\det T^{(\nu)}=1.
\end{equation}

\section{Transmission coefficient, reflection coefficient and resonance poles}
\label{sec-numerical}
\setcounter{equation}{0}

To summarize the derivation in the previous section, we have for a Cantor set of the $\nu$th generation, the relation
\begin{equation}\label{eq4000}
\left(
\begin{array}{c}
A_\mathrm{Rout} \\
A_\mathrm{Rin}
\end{array}
\right)
=T^{(\nu)}(\zeta)
\left(
\begin{array}{c}
A_\mathrm{Lin} \\
A_\mathrm{Lout}
\end{array}
\right),
\end{equation}
where $A_\mathrm{Lin}$ and $A_\mathrm{Lout}$ are the amplitudes of the incoming and outgoing waves on the left of the Cantor set, respectively, and $A_\mathrm{Rin}$ and $A_\mathrm{Rout}$ are those on the right, respectively.
(We here included phase factors such as $\E^{\I k_1 X \sigma_z}$ in the wave amplitudes.)
In the present section, we give the transmission and reflection coefficients as well as the resonance condition in terms of the elements of the transfer matrix $T^{(\nu)}$.
\begin{figure}
\begin{center}
\includegraphics[height=0.57\textheight]{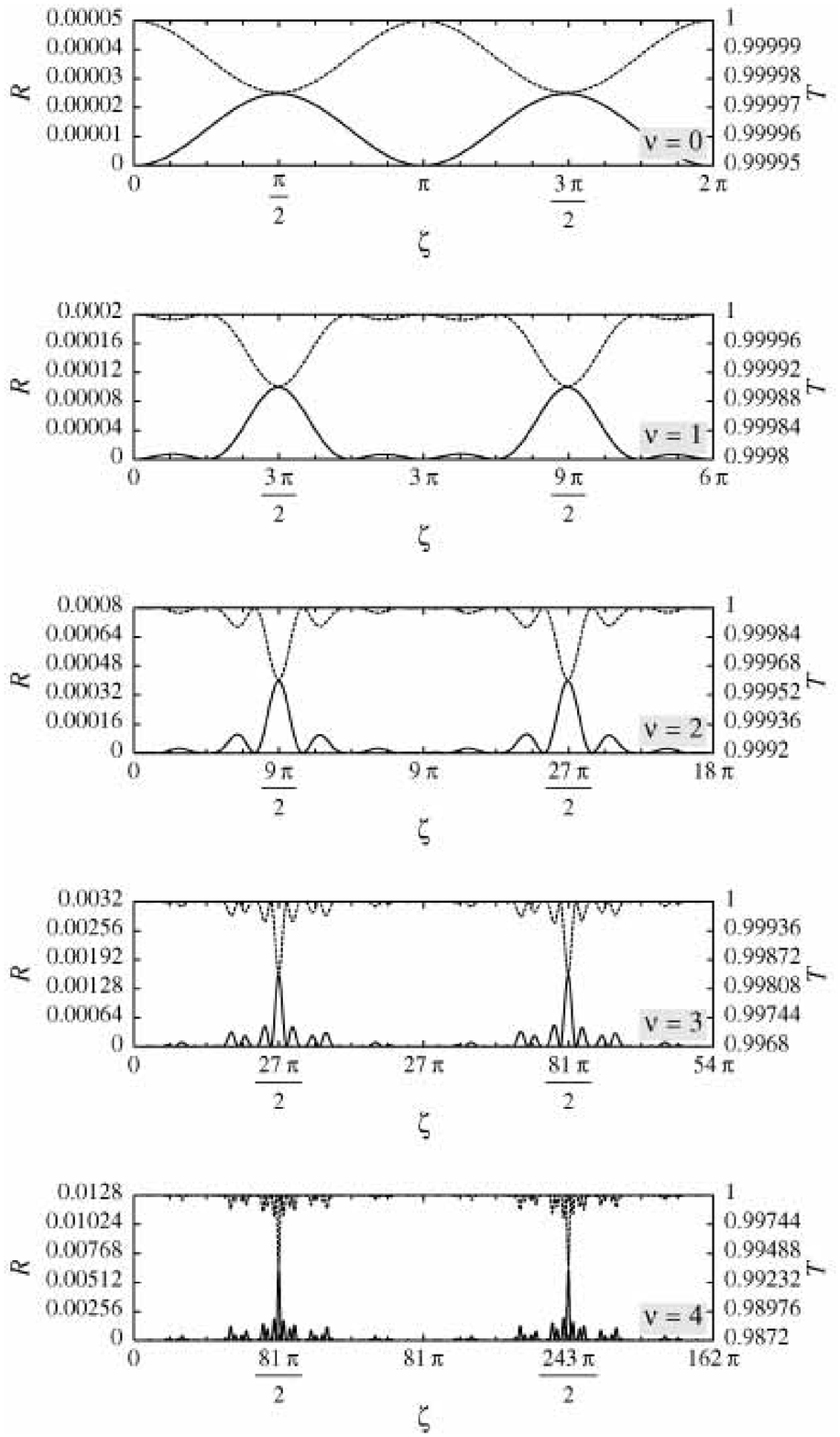}

(a)
\end{center}
\caption{The reflection coefficient $R$ (the solid curves) and the transmission coefficient $T$ (the broken curves) are plotted for $0\leq\nu\leq4$ (from the top down to the bottom) and for (a) $\varepsilon=1.01$, (c) $\varepsilon=1.5$ and (e) $\varepsilon=10$, with the resonance poles $\zeta=\zeta_\mathrm{r}\equiv\xi_\mathrm{r}+\I\eta_\mathrm{r}$ shown for (b) $\varepsilon=1.01$, (d) $\varepsilon=1.5$ and (f) $\varepsilon=10$.
On the panel (a), refer to the left axis for the reflection coefficient and to the right axis for the transmission coefficient.
On the panels (b), (d) and (e), the dots indicate the resonance poles.
Note that the horizontal axis of each panel is scaled by the factor $3^\nu$.}
\label{figTandR}
\end{figure}
\addtocounter{figure}{-1}
\begin{figure}
\begin{center}
\includegraphics[height=0.57\textheight]{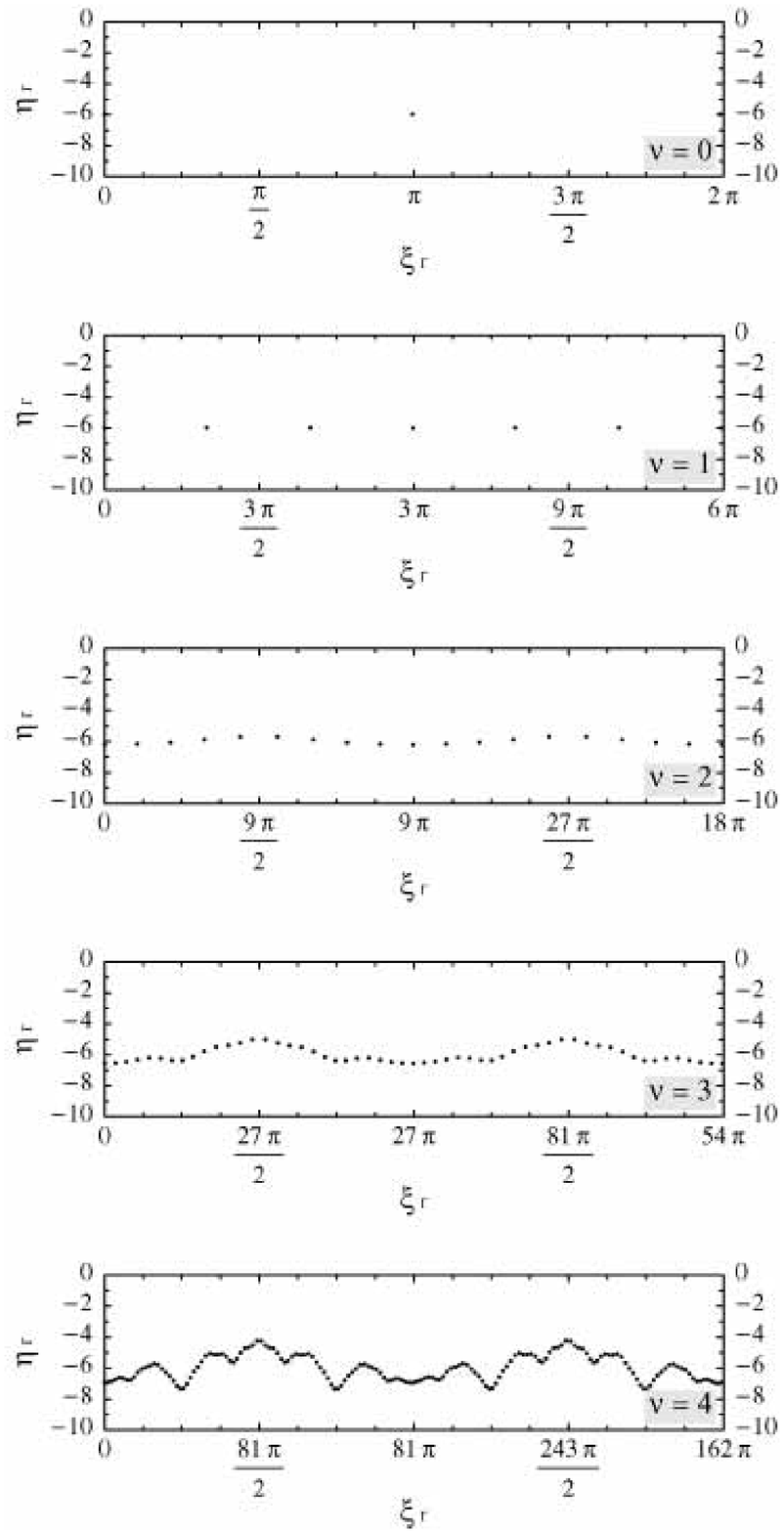}

(b)
\end{center}
\caption{\textit{Continued.}}
\end{figure}
\addtocounter{figure}{-1}
\begin{figure}
\begin{center}
\includegraphics[height=0.57\textheight]{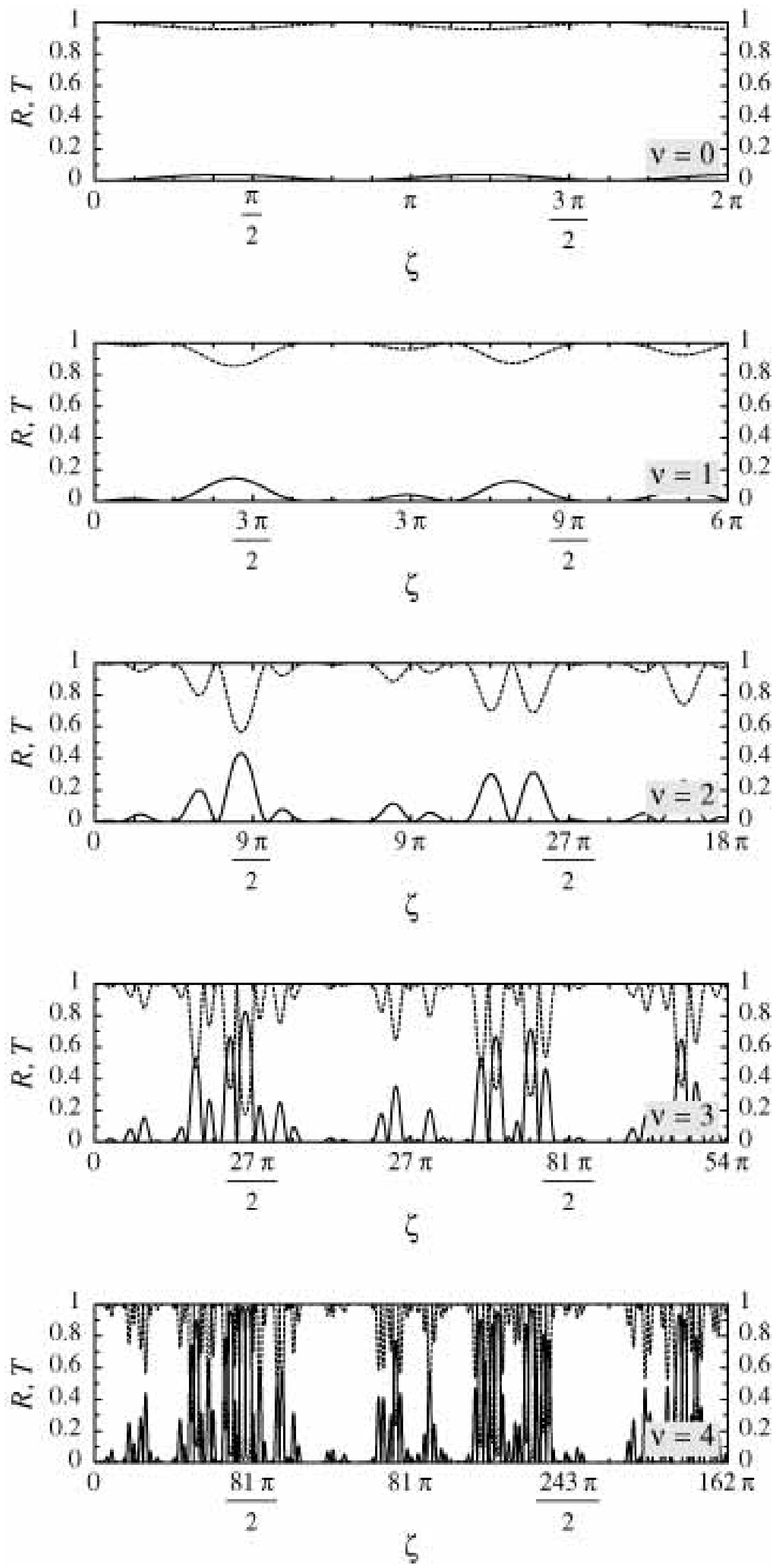}

(c)
\end{center}
\caption{\textit{Continued.}}
\end{figure}
\addtocounter{figure}{-1}
\begin{figure}
\begin{center}
\includegraphics[height=0.57\textheight]{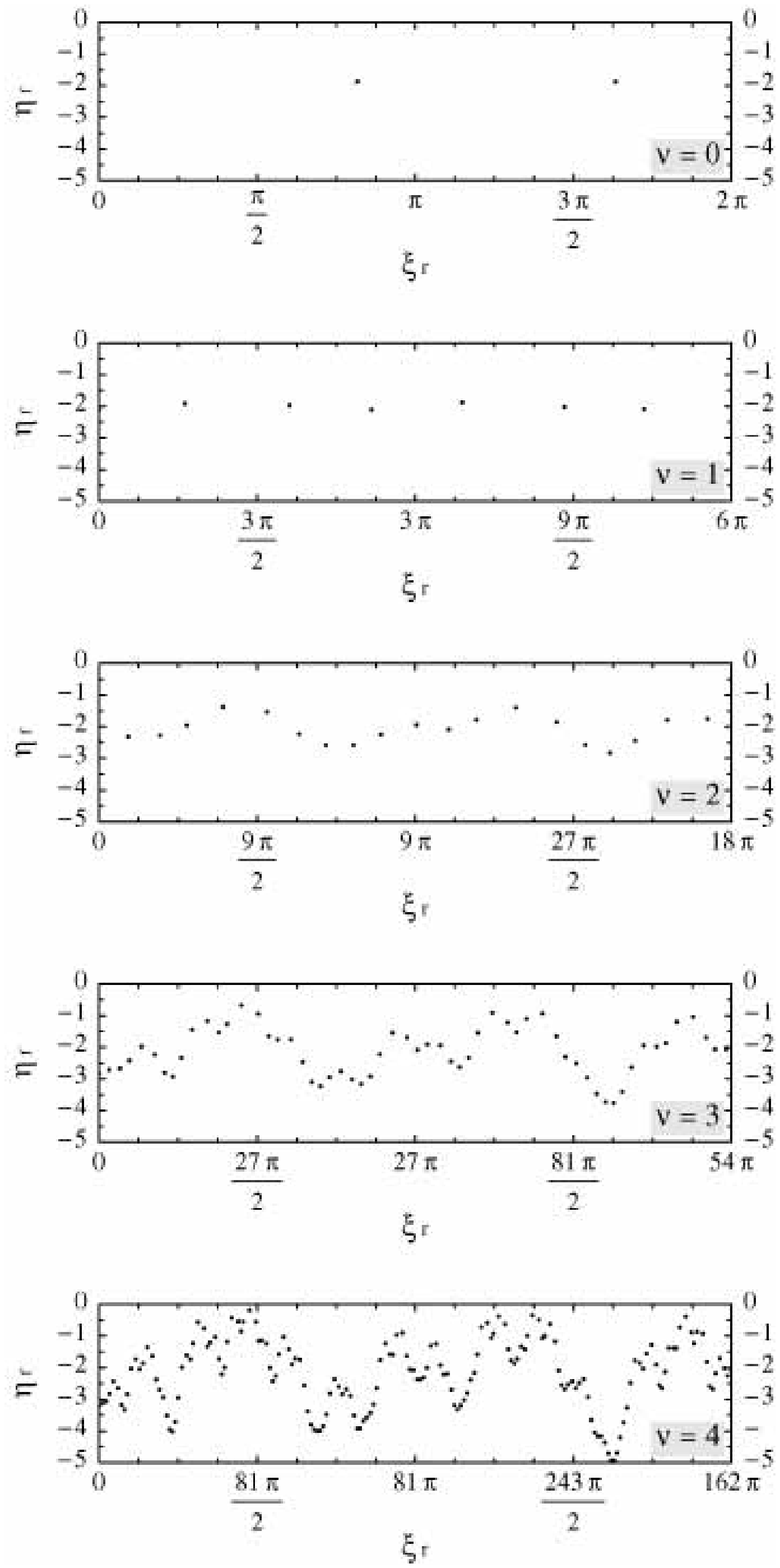}

(d)
\end{center}
\caption{\textit{Continued.}}
\end{figure}
\addtocounter{figure}{-1}
\begin{figure}
\begin{center}
\includegraphics[height=0.57\textheight]{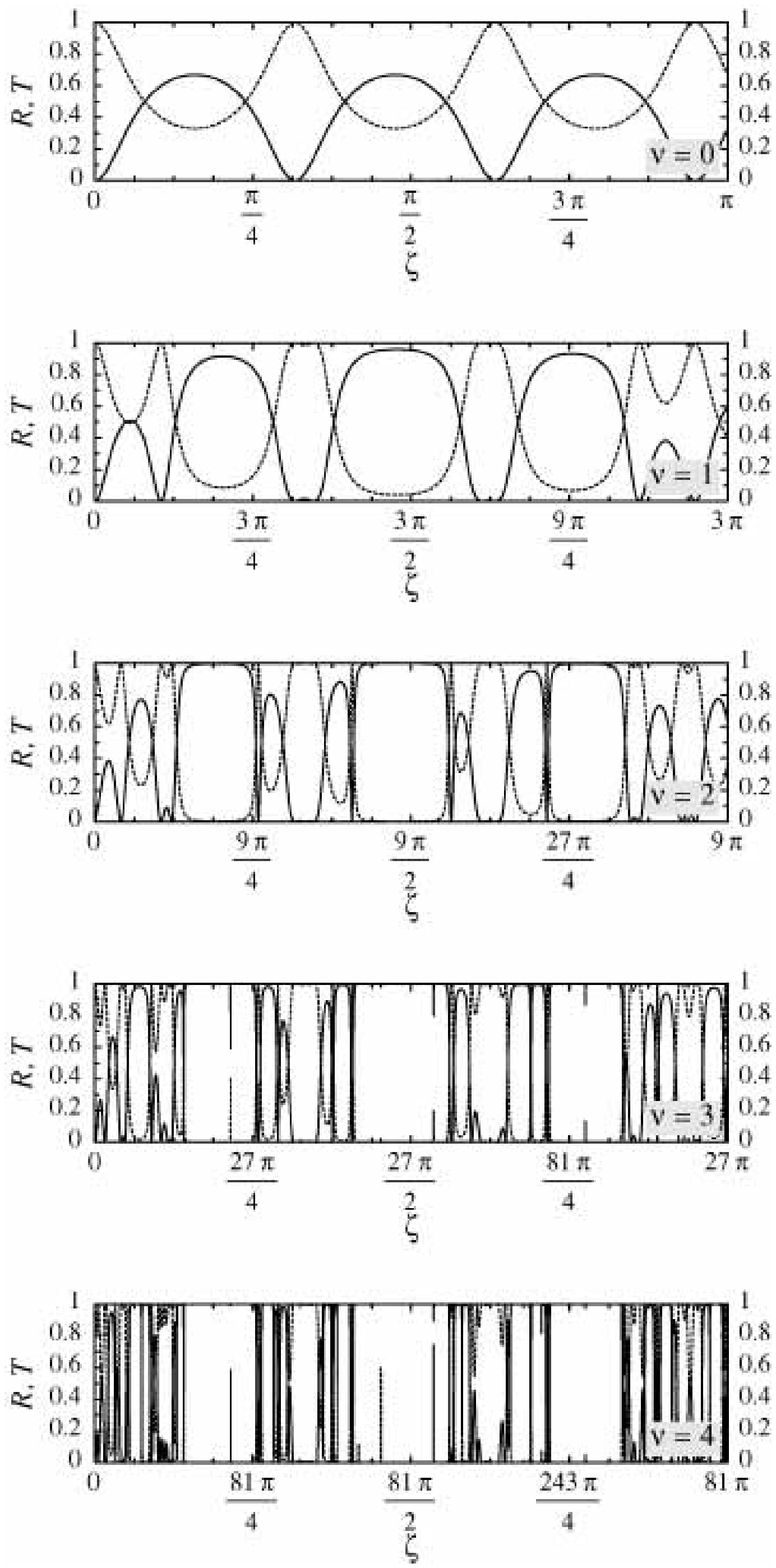}

(e)
\end{center}
\caption{\textit{Continued.}}
\end{figure}
\addtocounter{figure}{-1}
\begin{figure}
\begin{center}
\includegraphics[height=0.57\textheight]{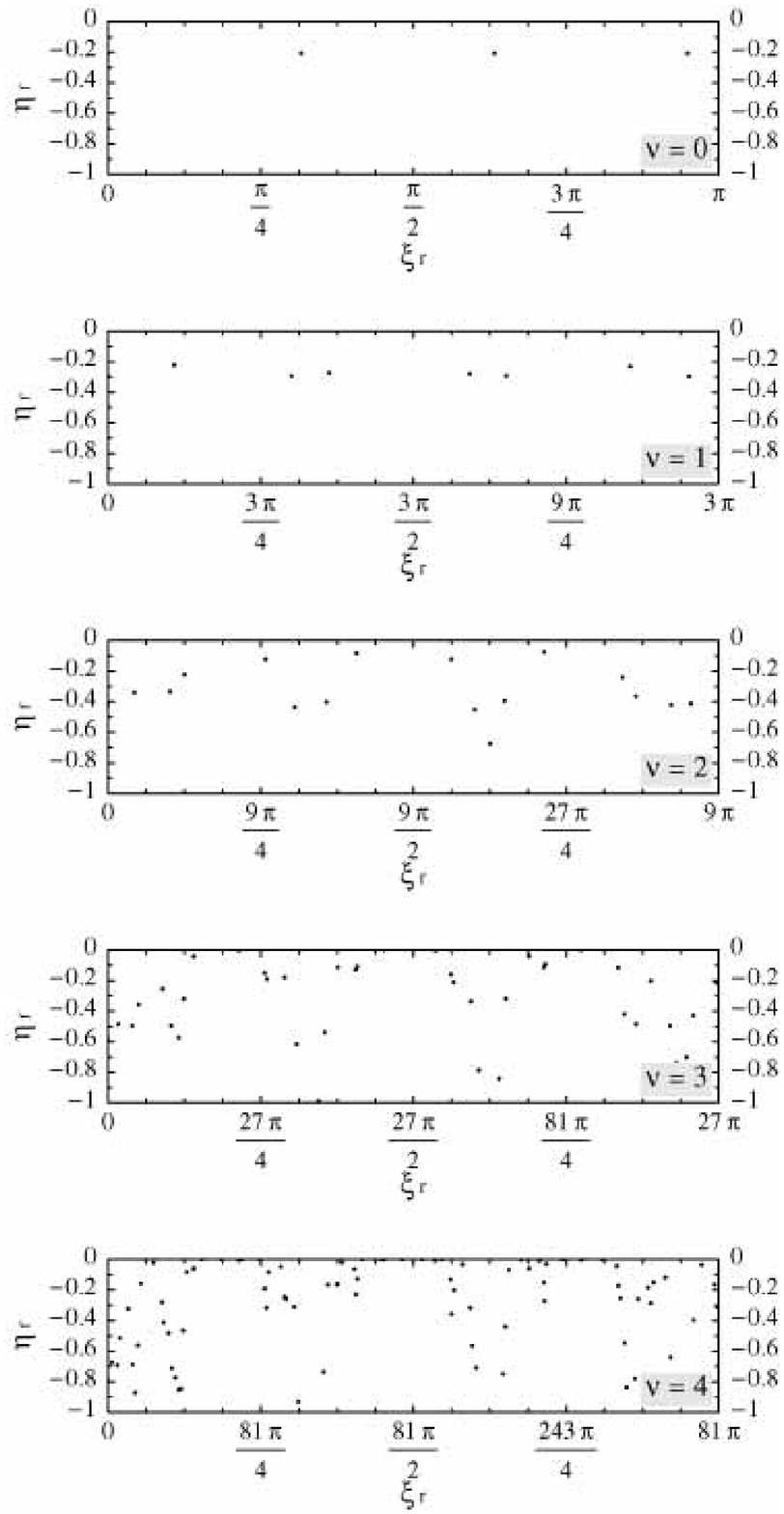}

(f)
\end{center}
\caption{\textit{Continued.}}
\end{figure}

Rearranging the transfer matrix in the form of a scattering matrix, we have
\begin{eqnarray}\label{eq4010}
\lefteqn{
\left(
\begin{array}{c}
A_\mathrm{Lout} \\
A_\mathrm{Rout}
\end{array}
\right)
}
\nonumber\\
&&
=
\frac{1}{T_{22}}
\left(
\begin{array}{cc}
-T_{21} & 1 \\
T_{11}T_{22}-T_{12}T_{21} & T_{12}
\end{array}
\right)
\left(
\begin{array}{c}
A_\mathrm{Lin} \\
A_\mathrm{Rin}
\end{array}
\right)
\nonumber\\
&&
=
\frac{1}{T_{22}}
\left(\begin{array}{cc}
-T_{12} & 1 \\
1 & T_{12}
\end{array}\right)
\left(
\begin{array}{c}
A_\mathrm{Lin} \\
A_\mathrm{Rin}
\end{array}
\right),
\end{eqnarray}
where we abbreviated $\left(T^{(\nu)}(\zeta)\right)_{ij}$ to $T_{ij}$ for simplicity.
(We used the symmetries~(\ref{eq330}) and~(\ref{eq340}) in moving from the second line to the third line of Eq.~(\ref{eq4010}).)
We thereby obtain the transmission and reflection coefficients as
\begin{equation}\label{eq4020}
T=\left| \frac{1}{T_{22}} \right|^2
\qquad\mbox{and}\qquad
R=\left| \frac{T_{12}}{T_{22}} \right|^2
\end{equation}
Here, the flux conservation $T+R=1$ is followed by the identity
\begin{equation}\label{eq4025}
\left|T_{12}\right|^2+1=\left|T_{22}\right|^2
\qquad\mbox{for $\zeta\in\mathbb{R}$.}
\end{equation}
We computed the transmission and reflection coefficients according to Eq.~(\ref{eq4020}) and plotted them in Fig.~\ref{figTandR}~(a),~(c) and~(e).
Notable is the wide regions of nearly total reflection ($R\sim1$) and sharp peaks of the transmission coefficient in the panel~(e).

We can relate the peaks and the dips of the transmission and reflection coefficients to resonance poles.
There are various ways of defining the resonant state.
One way of defining the resonant state is to seek a pole of the transmission and reflection coefficients~(\ref{eq4020}).
The resonance condition is thus given by
\begin{equation}\label{eq4030}
T_{22}=0.
\end{equation}
Another way of defining the resonant state is to put the incoming wave amplitudes to zero \cite{Siegert,Landau,Hatano};
in other words, the resonant state as a solution of the stationary wave equation is a function with outgoing waves only.
(This is sometimes called the Siegert condition.)
We obtain the solution by putting $A_\mathrm{Rin}=A_\mathrm{Lin}=0$ in Eq.~(\ref{eq4000}).
This also yields the condition~(\ref{eq4030}).

The condition has resonance solutions
\begin{equation}\label{eq4090}
\zeta=\zeta_\mathrm{r}\equiv\xi_\mathrm{r}+\I\eta_\mathrm{r}
\end{equation}
in the complex $\zeta$ plane.
(Note, however, that the imaginary part of the resonant eigenvalue is in fact negative: $\eta_\mathrm{r}<0$.\cite{Landau,Hatano}
A solution with a positive imaginary part would be a function with incoming waves only.)
The resonant frequency $\omega_\mathrm{r}$ is given by
\begin{equation}\label{eq4100}
\omega_\mathrm{r}=\frac{c_1}{L}\xi_\mathrm{r}
\end{equation}
and the resonant lifetime is given by
\begin{equation}\label{eq4110}
\tau_\mathrm{r}=\frac{L}{c_1}\frac{1}{\left|\eta_\mathrm{r}\right|}.
\end{equation}
We plot numerically obtained resonance solutions in Fig.~\ref{figTandR}~(b), (d) and (f);
each dot indicates the divergence of $1/\left|T_{22}\right|$, or a solution of the resonance condition~(\ref{eq4030}).

For $\varepsilon=10$ and $\nu=4$, we found resonance solutions with quite long lifetimes.
We list in Table~\ref{tableres} the solutions whose imaginary parts are less than $1.0\times10^{-5}$.
\begin{table}
\caption{Some of the resonance poles $\zeta_\mathrm{r}=\xi_\mathrm{r}+\I\eta_\mathrm{r}$ for $\varepsilon=10$ and $\nu=4$ with particularly long lifetimes.
We put $L=10$cm and $c_1=2.9979\times10^8$m for calculation of the resonant frequency $\omega_\mathrm{r}$ and the resonant lifetime $\tau_\mathrm{r}$.}
\label{tableres}
\vspace{\baselineskip}
\begin{center}
\begin{tabular}{cllll}
\hline
No. & \multicolumn{1}{c}{$\xi_\mathrm{r}$} & \multicolumn{1}{c}{$\eta_\mathrm{r}$} & \multicolumn{1}{c}{$\omega_\mathrm{r}$[THz]} & \multicolumn{1}{c}{$\tau_\mathrm{r}$[ms]} \\
\hline\hline
(a) & $47.2946$ & $-2.34999\times10^{-6}$ & $0.142$ & $0.142$ \\
\hline
(b) & $122.427$ & $-7.68867\times10^{-8}$ & $0.367$ & $4.34$ \\
\hline
(c) & $130.988$ & $-5.91115\times10^{-7}$ & $0.393$ & $0.564$ \\
\hline
(d) & $189.149$ & $-8.38728\times10^{-6}$ & $0.567$ & $0.0398$ \\
\hline
(e) & $292.234$ & $-6.21849\times10^{-6}$ & $0.876$ & $0.0536$ \\
\hline
(f) & $358.708$ & $-1.43720\times10^{-6}$ & $1.08$ & $0.232$ \\
\hline
(g) & $375.795$ & $-5.26455\times10^{-7}$ & $1.13$ & $0.634$ \\
\hline
(h) & $442.515$ & $-2.77803\times10^{-6}$ & $1.33$ & $0.120$ \\
\hline
(i) & $612.142$ & $-5.56439\times10^{-7}$ & $1.84$ & $0.599$ \\
\hline
(j) & $669.973$ & $-3.70414\times10^{-6}$ & $2.01$ & $0.0901$ \\
\hline
(k) & $687.429$ & $-6.88295\times10^{-6}$ & $2.06$ & $0.0485$ \\
\hline
(l) & $839.574$ & $-7.57655\times10^{-6}$ & $2.52$ & $0.0440$ \\
\hline
(m) & $857.027$ & $-3.24869\times10^{-6}$ & $2.57$ & $0.103$ \\
\hline
(n) & $914.846$ & $-7.07779\times10^{-7}$ & $2.74$ & $0.471$ \\
\hline
\end{tabular}
\end{center}
\end{table}
(We obtained the resonance solutions as follows: we first wrote down the resonance condition~(\ref{eq4030}) with Mathematica explicitly; we then solved the equation on the complex wave-number plane numerically by the Newton-Lapson method.
We can easily choose the initial guess for the Newton-Lapson method, since the distance between the real parts of the neighboring resonance poles is roughly equal for all poles.)

At the resonance solutions on the complex $\zeta$ plane, the transmission and reflection coefficients~(\ref{eq4020}) diverge.
The tail of the divergence forms structures on the real $\zeta$ axis as is exemplified in Fig.~\ref{figTandR}.
On the real axis near the resonance solutions, the transmission coefficient has a peak: $T\sim1$;
see, for example, the resonance pole at $(\xi_\mathrm{r},\eta_\mathrm{r})\simeq(5\pi/16,-0.2)$ for $\nu=0$ in Fig.~\ref{figTandR}~(f) and the corresponding peak of the transmission coefficient $T$ for $\nu=0$ in Fig.~\ref{figTandR}~(e).
Because of the conservation law, the reflection coefficient then must have a dip: $R\sim0$.

Finally, we show the spatial distribution of the field $A(x)$.
We can compute the field amplitude inside the Cantor set by expressing the transfer matrix $T^{(\nu)}$ with the elementary transfer matrices $T_1$ and $T_2$ explicitly.
We apply the elementary transfer matrices one by one to the amplitude vector and obtain the field amplitudes inside the Cantor set.

\begin{figure}
\begin{center}
\includegraphics[height=0.85\textheight]{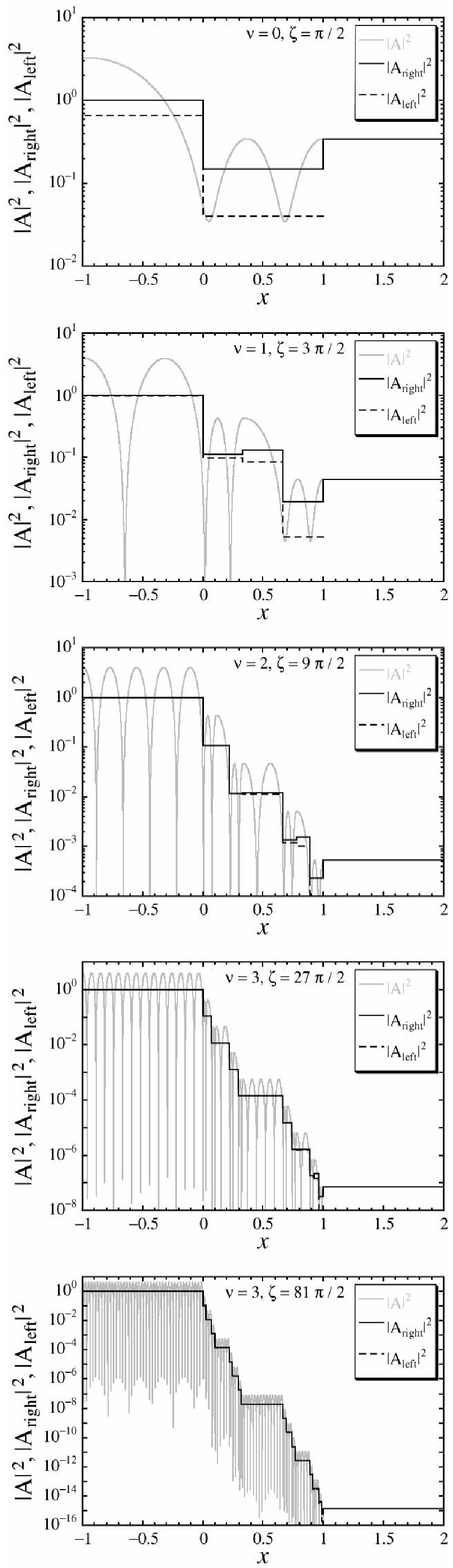}
\end{center}
\caption{Semi-logarithmic plots of the field distribution $A(x)$ for $\varepsilon=10$ and $0\leq\nu\leq 4$ (from the top down to the bottom) at $\zeta=k_1L=3^\nu\pi/2$.
(The condition $\zeta=3^\nu m\pi/2$ shown in Fig.~\ref{cavity} is on resonance only for $\varepsilon\sim1$.
It is off resonance for $\varepsilon=10$.)
Each solid line indicates the magnitude of the right-going wave, $\left|A_\mathrm{right}\right|^2$, each broken line indicates the magnitude of the left-going wave, $\left|A_\mathrm{left}\right|^2$, and each gray line indicates the magnitude of the total standing wave, $\left|A(x)\right|^2=\left| A_\mathrm{right}\E^{\I k_ix}+A_\mathrm{left}\E^{-\I k_i x}\right|^2$ in the medium $i(=1,2)$.}
\label{field}
\end{figure}

Figure~\ref{field} shows the field distribution for $\varepsilon=10$ and, in particular, for the cases that the reflection coefficient is close to unity.
In this figure, $A_\mathrm{right}$ and $A_\mathrm{left}$ denote the amplitudes of the right-going component and the left-going component, respectively;
that is, the total field is given by
\begin{equation}\label{eq4120}
A(x)=A_\mathrm{right}\E^{\I k_i x}+A_\mathrm{left}\E^{-\I k_i x}
\end{equation}
in the medium $i(=1,2)$.
We see that the incoming wave from the left is bounced almost completely and barely transmits to the right.

\begin{figure}
\begin{center}
\includegraphics[width=0.8\columnwidth]{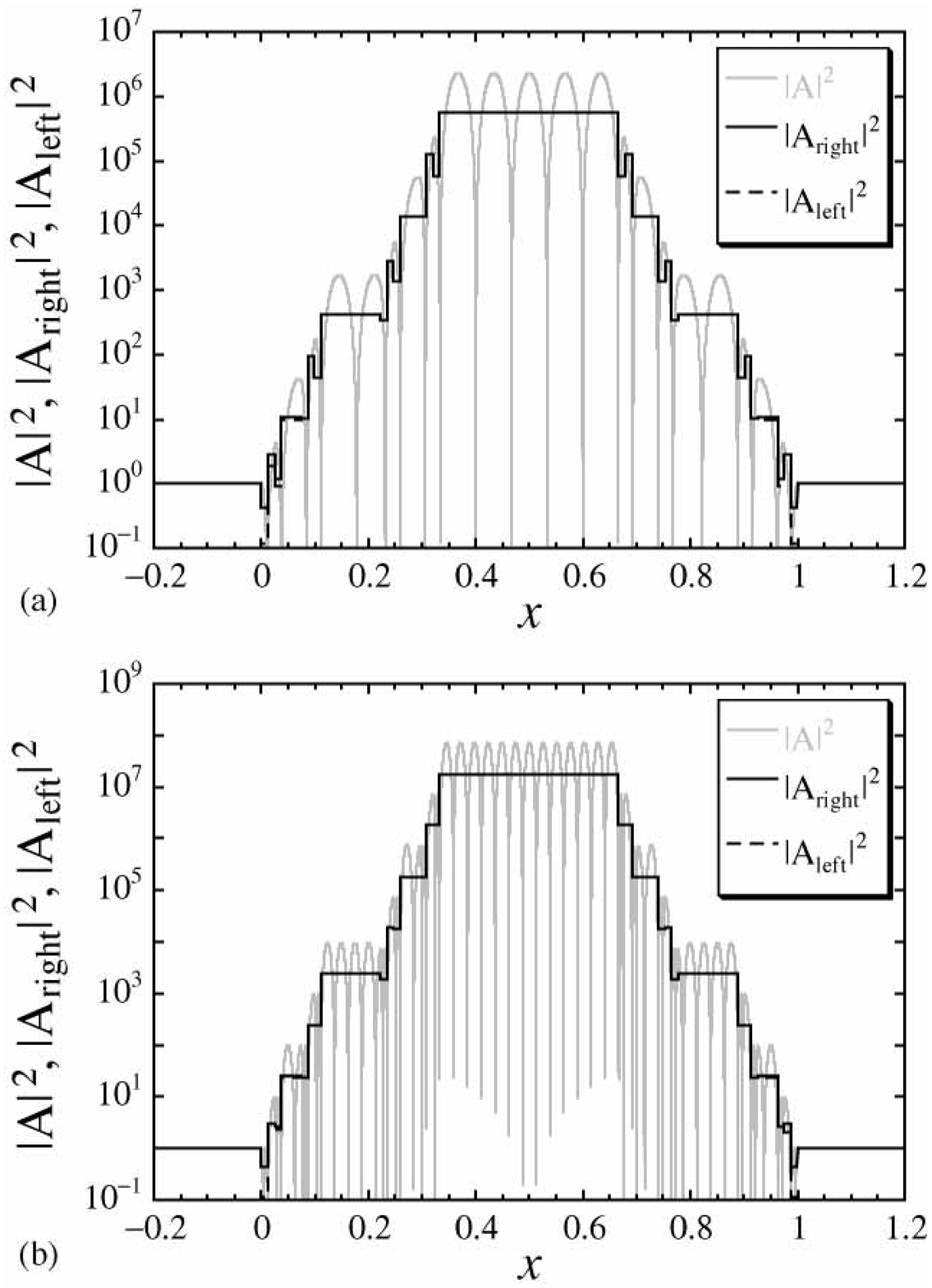}
\end{center}
\caption{Semi-logarithmic plots of the field distribution $A(x)$ for $\varepsilon=10$ and $\nu=4$ at (a) $\zeta=k_1L=47.29458732802431$ and (b) $\zeta=k_1L=122.4274149967578$.
Each solid line indicates the magnitude of the right-going wave, $\left|A_\mathrm{right}\right|^2$, each broken line indicates the magnitude of the left-going wave, $\left|A_\mathrm{left}\right|^2$, and each gray line indicates the magnitude of the total standing wave, $\left|A(x)\right|^2=\left| A_\mathrm{right}\E^{\I k_ix}+A_\mathrm{left}\E^{-\I k_i x}\right|^2$ in the medium $i(=1,2)$.}
\label{fieldres}
\end{figure}

On the other hand, Fig.~\ref{fieldres} shows the field distribution in the cases close to the resonance poles given in Table~\ref{tableres}~(a) and (b).
We see prominent localization of the field inside the Cantor set.

Let us show that the enhancement of the field inside the Cantor set is of the order of the inverse of the imaginary part of the resonance pole.
Suppose that the incident wave $A_\mathrm{Lin}$ from the left of the Cantor set of the $\nu$th generation causes a reflection wave $A_\mathrm{Lout}$ and a transmission wave $A_\mathrm{Rout}$.
Then we have the relation~(\ref{eq4000}), or
\begin{equation}\label{eq7000}
\left(
\begin{array}{c}
A_\mathrm{Rout} \\
0
\end{array}
\right)
=T^{(\nu)}(\zeta)
\left(
\begin{array}{c}
A_\mathrm{Lin} \\
A_\mathrm{Lout}
\end{array}
\right).
\end{equation}
Near the resonance but on the real $\zeta$ axis, the transmission coefficient is close to unity, which means $\left|A_\mathrm{Rout}\right|\simeq\left|A_\mathrm{Lin}\right|$ and $A_\mathrm{Lout}\simeq0$.
We are now interested in the right-going field $A_\mathrm{right}$ and the left-going field $A_\mathrm{left}$ inside the cavity $L/3<x<2L/3$.
This is related to $A_\mathrm{Rout}$ as
\begin{equation}\label{eq7010}
\left(
\begin{array}{c}
A_\mathrm{Rout} \\
0
\end{array}
\right)
=T^{(\nu-1)}\left(\frac{\zeta}{3}\right)
\left(
\begin{array}{c}
A_\mathrm{right} \\
A_\mathrm{left}
\end{array}
\right).
\end{equation}
Let us for the moment use the following symbols for the sake of simplicity:
\begin{eqnarray}\label{eq7020}
T^{(\nu)}(\zeta)&=&
\left(\begin{array}{cc}
T_{11} & T_{12} \\
T_{21} & T_{22}
\end{array}\right)
\\
T^{(\nu-1)}\left(\frac{\zeta}{3}\right)&=&
\left(\begin{array}{cc}
t_{11} & t_{12} \\
t_{21} & t_{22}
\end{array}\right).
\end{eqnarray}
Equation~(\ref{eq7010}) takes the form of scattering of an incoming wave $A_\mathrm{right}$ due to the part $2L/3<x<L$ of the medium 2.
Hence the transmission coefficient due to the part $2L/3<x<L$ gives the enhancement of the wave amplitude as follows:
\begin{equation}\label{eq7025}
\left|A_\mathrm{Rout}\right|^2=
\left|\frac{1}{t_{22}}\right|^2\left|A_\mathrm{right}\right|^2,
\end{equation}
or
\begin{equation}\label{eq7026}
\frac{\left|A_\mathrm{right}\right|^2}{\left|A_\mathrm{Lin}\right|^2}
\simeq
\frac{\left|A_\mathrm{right}\right|^2}{\left|A_\mathrm{Rout}\right|^2}
=\left|t_{22}\right|^2.
\end{equation}
Because of the spatial symmetry at the resonance, we also have $\left|A_\mathrm{right}\right|\simeq\left|A_\mathrm{left}\right|$.
This is indeed confirmed by considering a relation from Eq.~(\ref{eq7010}),
\begin{equation}\label{eq7028}
\left|A_\mathrm{left}\right|^2=
\left|\frac{t_{12}}{t_{22}}\right|^2\left|A_\mathrm{right}\right|^2
\simeq\left|A_\mathrm{right}\right|^2,
\end{equation}
where we used the fact that $\left|t_{21}\right|\simeq\left|t_{22}\right|$ near the resonance as will be seen below in Eq.~(\ref{eq7060}).
Hence the maximum amplitude of the standing wave inside the cavity is
\begin{equation}\label{eq7027}
M\equiv
\frac{{\displaystyle \max_x}\left|A_\mathrm{right}\E^{\I k_1 x}+A_\mathrm{left}\E^{-\I k_1 x}\right|^2}%
{\left|A_\mathrm{Lin}\right|^2}
\simeq
4\left|t_{22}\right|^2
\end{equation}
with its average half of Eq.~(\ref{eq7027}).

Now we relate the quantity $\left|t_{22}\right|^2$ to the transfer matrix $T^{(\nu)}$.
Because of the recursion relation~(\ref{eq320}), the resonance condition~(\ref{eq4030}) reads
\begin{equation}\label{eq7030}
T_{22}=\E^{\I\zeta/3}\left.t_{12}\right.^2+\E^{-\I\zeta/3}\left.t_{22}\right.^2=0
\end{equation}
at $\zeta=\zeta_\mathrm{r}=\xi_\mathrm{r}+\I \eta_\mathrm{r}$, where we used the fact $t_{12}=t_{21}$ on the basis of Eq.~(\ref{eq330}).
We transform Eq.~(\ref{eq7030}) into
\begin{equation}
\label{eq7050}
\left|t_{12}\left(\frac{\zeta_\mathrm{r}}{3}\right)\right|^2
=\E^{2\eta_\mathrm{r}/3}\left|t_{22}\left(\frac{\zeta_\mathrm{r}}{3}\right)\right|^2.
\end{equation}
For a very small $\left|\eta_\mathrm{r}\right|=-\eta_\mathrm{r}$, we can expand the both sides in the form
\begin{eqnarray}\label{eq7060}
\lefteqn{
\left|t_{12}\left(\frac{\xi_\mathrm{r}}{3}\right)\right|^2+\mathrm{O}(\eta_\mathrm{r})
}
\nonumber\\
&&=\left(1+\frac{2}{3}\eta_\mathrm{r}\right)
\left(\left|t_{22}\left(\frac{\xi_\mathrm{r}}{3}\right)\right|^2+\mathrm{O}(\eta_\mathrm{r})\right).
\end{eqnarray}
Using the relation~(\ref{eq4025}), or $\left|t_{12}(\xi_\mathrm{r}/3)\right|^2+1=\left|t_{22}(\xi_\mathrm{r}/3)\right|^2$, we can reduce Eq.~(\ref{eq7060}) into
\begin{equation}\label{eq7070}
\left|t_{22}\left(\frac{\xi_\mathrm{r}}{3}\right)\right|^2
\simeq\frac{3}{2\left|\eta_\mathrm{r}\right|}.
\end{equation}
(Note again the fact that $\eta_\mathrm{r}<0$.\cite{Landau,Hatano})
The enhancement factor given in Eq.~(\ref{eq7027})  is thereby
\begin{equation}\label{eq7075}
M\simeq\frac{6}{\left|\eta_\mathrm{r}\right|}
\end{equation}
and the enhancement factor on average is $3/\left|\eta_\mathrm{r}\right|$.
This is indeed consistent with the numerical results in Fig.~\ref{fieldres};
the enhancement factor in Fig.~\ref{fieldres}~(a) is given by Table~\ref{tableres}~(a) as $M=2.5\times10^6$ and that in Fig.~\ref{fieldres}~(b) is given by Table~\ref{tableres}~(b) as $M=7.8\times10^7$.

Finally, we make a note for discussions in the final section \S\ref{sec-discuss}.
Notice that the field distribution in Fig.~\ref{fieldres} is almost symmetric, being independent of the direction of the incident wave;
although the incident wave, in fact, comes from the left in Fig.~\ref{fieldres}, the difference between the field strengths on the left end and the right end of the Cantor set is negligible.
This is a common feature of the resonance.
As we argued below Eq.~(\ref{eq4030}), we have only outgoing waves at the resonance, and hence the resonant wave function itself is either symmetric or anti-symmetric, depending on the parity of the state.
The field strengths on the left and the right are therefore equal.
This is strictly correct only at the resonance, or on the complex plane.
In the present case, however, the resonance pole is so close to the real axis that the field distribution is almost symmetric at the resonant frequency on the real axis, as we showed in Eqs.~(\ref{eq7028}) and~(\ref{eq7060}).

We stress here that this happens for the three-dimensional resonance as well.
For the Menger sponge embedded in three spatial dimensions, each resonant wave function corresponds to a irreducible representation of the cubic point group O$_\mathrm{h}$.
Some of the resonant states have outgoing waves of the same amplitude in all six directions.
At the resonant frequency close to such resonant states but on the real axis, the scattered wave has a strong amplitude even in the direction perpendicular to the incident wave vector.

\section{Renormalization}
\label{sec-renom}
\setcounter{equation}{0}

In this section, we introduce a renormalization procedure of the transfer matrix for the Cantor set.
We remind the readers of the definition of the transfer matrix $T_2$ in Eq.~(\ref{eq280}):
\begin{equation}\label{eq1045}
T_2(\zeta)\equiv
\E^{-\I\phi\sigma_y/2}\E^{\I \zeta n \sigma_z}\E^{\I\phi\sigma_y/2}
=\E^{\I \zeta n (\sigma_z \cos\phi+\sigma_x \sin\phi)}.
\end{equation}
The exponential operators of $\sigma_y$ on the left-hand side represents the effects of the boundaries at $x=X$ and $x=Y$.
We spread the effect of the reflection at the boundaries over the entire medium~2 in $X\leq x\leq Y$.
In other words, we renormalized the boundary effect into a uniform medium and hence had the single exponential operator on the right-hand side, which takes the form of propagation in a uniform medium, similar to the form~(\ref{eq300}).
The matrix $\sigma_z$ in the exponent of the right-hand side represents the straight propagation, whereas the matrix $\sigma_x$ represents the reflection inside the medium~2.

We are now going to exploit the above idea and represent the transfer matrix for the Cantor set of any generation in the form of the transfer matrix for a uniform medium.
In other words, we renormalize the fractal structure into an effective uniform medium.

We first prove that the transfer matrix~$T^{(\nu)}$ for an arbitrary $\nu$ is represented in the form
\begin{equation}\label{eq1050}
T^{(\nu)}(\zeta)=\E^{\I\zeta(s^{(\nu)}\sigma_z-\I r^{(\nu)}\sigma_x)}
\end{equation}
with appropriate coefficients $s^{(\nu)}$ and $r^{(\nu)}$.
First, the symmetry of the transfer matrix, Eq.~(\ref{eq330}), gives the symmetry of its exponent as
\begin{equation}\label{eq1060}
{}^{\mathrm{t}}\log T^{(1)}(\zeta)=\log{}^{\mathrm{t}} T^{(1)}(\zeta)
=\log T^{(1)}(\zeta).
\end{equation}
Next, the exponents of the transfer matrices $T_1$ and $T_2$ are traceless.
Therefore we have
\begin{eqnarray}\label{eq1070}
\lefteqn{
\mathrm{Tr}\log T^{(1)}(\zeta)=\log\det T^{(1)}(\zeta)
}
\nonumber\\
&=&\log\det T_1\left(\frac{\zeta}{3}\right)
+2\log\det T_2\left(\frac{\zeta}{3}\right)
\nonumber\\
&=&\mathrm{Tr}\log T_1\left(\frac{\zeta}{3}\right)
+2\mathrm{Tr}\log T_2\left(\frac{\zeta}{3}\right)
=0.
\end{eqnarray}
Equations~(\ref{eq1060}) and~(\ref{eq1070}) indicate that the exponent of the matrix $T^{(1)}$ is symmetric and traceless.
We can similarly show from the recursion formula~(\ref{eq320}) that the exponents of all transfer matrices $T^{(\nu)}$ are symmetric and traceless.
An arbitrary symmetric traceless $2\times2$ matrix is expressed by a linear combination of $\sigma_z$ and $\sigma_x$.
Hence we have Eq.~(\ref{eq1050}).

Next, we prove that the coefficients $s^{(\nu)}$ and $r^{(\nu)}$ are real for real $\zeta$.
We first show it for $\nu=0$, or for $T_2$ defined by Eq.~(\ref{eq280}).
In this case, we have
\begin{equation}
s^{(0)}=n\cos\phi
\quad\mbox{and}\quad
r^{(0)}=\I n\sin\phi.
\end{equation}
Equations~(\ref{eq285}) and~(\ref{eq286}) indicate that $s^{(0)}$ and $r^{(0)}$ are both real (assuming that the dielectric constants are real).
Next, we derive that $s^{(\nu+1)}$ and $r^{(\nu+1)}$ are real from the assumption that $s^{(\nu)}$ and $r^{(\nu)}$ are real.
For the sake of the proof, we rotate in the recursion formula~(\ref{eq320}) the $x$ and $y$ axes of the spin space around the $z$ axis by 90 degrees and consider the identity
\begin{eqnarray}\label{eq1080}
\lefteqn{
\E^{\I\zeta(s^{(\nu+1)}\sigma_z-\I r^{(\nu+1)}\sigma_y)}
}
\nonumber\\
&&
=\E^{\I\zeta(s^{(\nu)}\sigma_z-\I r^{(\nu)}\sigma_y)/3}
\E^{\I\zeta\sigma_z/3}
\E^{\I\zeta(s^{(\nu)}\sigma_z-\I r^{(\nu)}\sigma_y)/3}.
\qquad
\end{eqnarray}
By taking the complex conjugate of Eq.~(\ref{eq1080}), we have
\begin{eqnarray}\label{eq1090}
\lefteqn{
\E^{-\I\zeta((s^{(\nu+1)})^\ast\sigma_z-\I (r^{(\nu+1)})^\ast\sigma_y)}
}
\nonumber\\
&&=\E^{-\I\zeta(s^{(\nu)}\sigma_z-\I r^{(\nu)}\sigma_y)/3}
\E^{-\I\zeta\sigma_z/3}
\nonumber\\
&&\qquad\qquad
\times
\E^{-\I\zeta(s^{(\nu)}\sigma_z-\I r^{(\nu)}\sigma_y)/3}
\end{eqnarray}
for $\zeta\in\mathbb{R}$.
(Note here the fact that $(\sigma_y)^\ast=-\sigma_y$.)
Hence we have
\begin{equation}\label{eq1100}
\E^{-\I\zeta((s^{(\nu+1)})^\ast\sigma_z-\I (r^{(\nu+1)})^\ast\sigma_y)}
=\E^{-\I\zeta(s^{(\nu+1)}\sigma_z-\I r^{(\nu+1)}\sigma_y)},
\end{equation}
or, by taking the logarithm,
\begin{equation}
(s^{(\nu+1)})^\ast\sigma_z-\I (r^{(\nu+1)})^\ast\sigma_y
=s^{(\nu+1)}\sigma_z-\I r^{(\nu+1)}\sigma_y.
\end{equation}
Because both sides of the equation are traceless $2\times2$ matrices, the expression with respect to the Pauli matrices must be unique.
Therefore we have $(s^{(\nu+1)})^\ast=s^{(\nu+1)}$ and $(r^{(\nu+1)})^\ast=r^{(\nu+1)}$ for $\zeta\in\mathbb{R}$.

To summarize this section, the transfer matrix given by Eq.~(\ref{eq320}) is expressed in the form
\begin{equation}\label{eq1105}
T^{(\nu)}(\zeta)=\E^{\I\zeta(s^{(\nu)}\sigma_z-\I r^{(\nu)}\sigma_x)}
\end{equation}
with real coefficients $s^{(\nu)}$ and $r^{(\nu)}$.
The former coefficient represents the (renormalized) straight propagation and the latter represents the (renormalized) reflection.

\section{Perturbation}
\label{sec-pert}
\setcounter{equation}{0}

We now would like to find fixed points $(s^{(\ast)}, r^{(\ast)})$ of the renormalization transformation
\begin{equation}\label{eq2005}
(s^{(\nu)},r^{(\nu)})\longrightarrow(s^{(\nu+1)},r^{(\nu+1)}).
\end{equation}
As a first step, we here argue the transformation perturbatively for $\varepsilon\equiv\varepsilon_2/\varepsilon_1\simeq1$.
For simplicity, we hereafter assume $\mu_1=\mu_2$ and hence $n'=n$.

Our perturbation parameter is
\begin{equation}\label{eq2010}
\lambda\equiv\frac{1}{2}\left(\frac{\varepsilon_2}{\varepsilon_1}-1\right)
=\frac{n^2-1}{2}.
\end{equation}
Equations~(\ref{eq285}) and~(\ref{eq286}) yield
\begin{eqnarray}\label{eq2020}
&&s^{(0)}=n\cos\phi
=\frac{1}{2}\left(\frac{\varepsilon_2}{\varepsilon_1}+1\right)=1+\lambda,
\qquad
\\
&&r^{(0)}=\I n\sin\phi
=\frac{1}{2}\left(\frac{\varepsilon_2}{\varepsilon_1}-1\right)=\lambda.
\qquad
\end{eqnarray}
Hence we have
\begin{equation}\label{eq2030}
T^{(0)}(\zeta)\equiv T_2(\zeta)=\E^{\I\zeta(A+\lambda C^{(0)})}
\end{equation}
with
\begin{equation}\label{eq2040}
A=\sigma_z
\qquad\mbox{and}\qquad
C^{(0)}=\sigma_z-\I\sigma_x.
\end{equation}

Now we analyze the recursion relation~(\ref{eq310}), or the identity
\begin{equation}\label{eq2050}
\E^{\I\zeta(A+\lambda C^{(0)})/3}\E^{\I\zeta A/3}\E^{\I\zeta(A+\lambda C^{(0)})/3}
=\E^{\I\zeta (A+\lambda C^{(1)} + \mathrm{O}(\lambda^2))}
\end{equation}
using M.~Suzuki's method of the \lq\lq quantum analysis."\cite{Suzuki1,Suzuki2}
The calculation in \ref{app-QA} gives Eq.~(\ref{eq3060}), or
\begin{equation}\label{eq3061}
C^{(1)}=\frac{2}{3}\sigma_z-\I F^{(1)}(\zeta)\sigma_x,
\end{equation}
where
\begin{eqnarray}\label{eq3080}
F^{(1)}(\zeta)&\equiv&\frac{\E^{\I\zeta}-\E^{\I\zeta/3}+\E^{-\I\zeta/3}-\E^{-\I\zeta}}{\E^{\I\zeta}-\E^{-\I\zeta}}
\\ \label{eq3081}
&=&1-\frac{\sin\frac{\zeta}{3}}{\sin\zeta}
=2\frac{\cos\frac{2\zeta}{3}\sin\frac{\zeta}{3}}{\sin\zeta}.
\end{eqnarray}
Comparing the both sides of
\begin{equation}\label{3085}
\E^{\I\zeta (A+\lambda C^{(1)})}
\simeq
\E^{\I\zeta(s^{(1)}\sigma_z-\I r^{(1)}\sigma_x)},
\end{equation}
we have the renormalization transformation
\begin{eqnarray}\label{eq3070}
\lefteqn{
s^{(0)}=1+\lambda
\quad\mbox{and}\quad
r^{(0)}=\lambda
\longrightarrow}
\nonumber\\
&&
s^{(1)}\simeq 1+\frac{2}{3}\lambda
\quad\mbox{and}\quad
r^{(1)}\simeq\lambda F^{(1)}(\zeta).
\end{eqnarray}
Similar calculation yields Eq.~(\ref{eq3130}), or
\begin{equation}\label{eq3140}
s^{(2)}\simeq 1+\left(\frac{2}{3}\right)^2\lambda
\quad\mbox{and}\quad
r^{(2)}\simeq\lambda F^{(2)}(\zeta)
\end{equation}
with
\begin{equation}\label{eq3150}
F^{(2)}(\zeta)\equiv F^{(1)}(\zeta)F^{(1)}\left(\frac{\zeta}{3}\right).
\end{equation}
We go forward similarly, arriving at
\begin{equation}\label{eq3160}
s^{(\nu)}\simeq 1+\left(\frac{2}{3}\right)^\nu\lambda
\quad\mbox{and}\quad
r^{(\nu)}(\zeta)\simeq\lambda F^{(\nu)}(\zeta)
\end{equation}
with
\begin{eqnarray}\label{eq3170}
F^{(\nu)}(\zeta)&=&F^{(1)}(\zeta)F^{(\nu-1)}\left(\frac{\zeta}{3}\right)
=\prod_{k=0}^{\nu-1}F^{(1)}\left(\frac{\zeta}{3^k}\right)
\nonumber\\
&=&2^\nu\frac{\sin\frac{\zeta}{3^\nu}}{\sin\zeta}\prod_{k=1}^\nu\cos\frac{2\zeta}{3^k}.
\end{eqnarray}
We also define
\begin{equation}\label{eq3171}
F^{(0)}(\zeta)\equiv1,
\end{equation}
which makes Eq.~(\ref{eq3160}) hold for $\nu=0$ as well.

\begin{figure}
\begin{center}
\includegraphics[width=0.8\columnwidth]{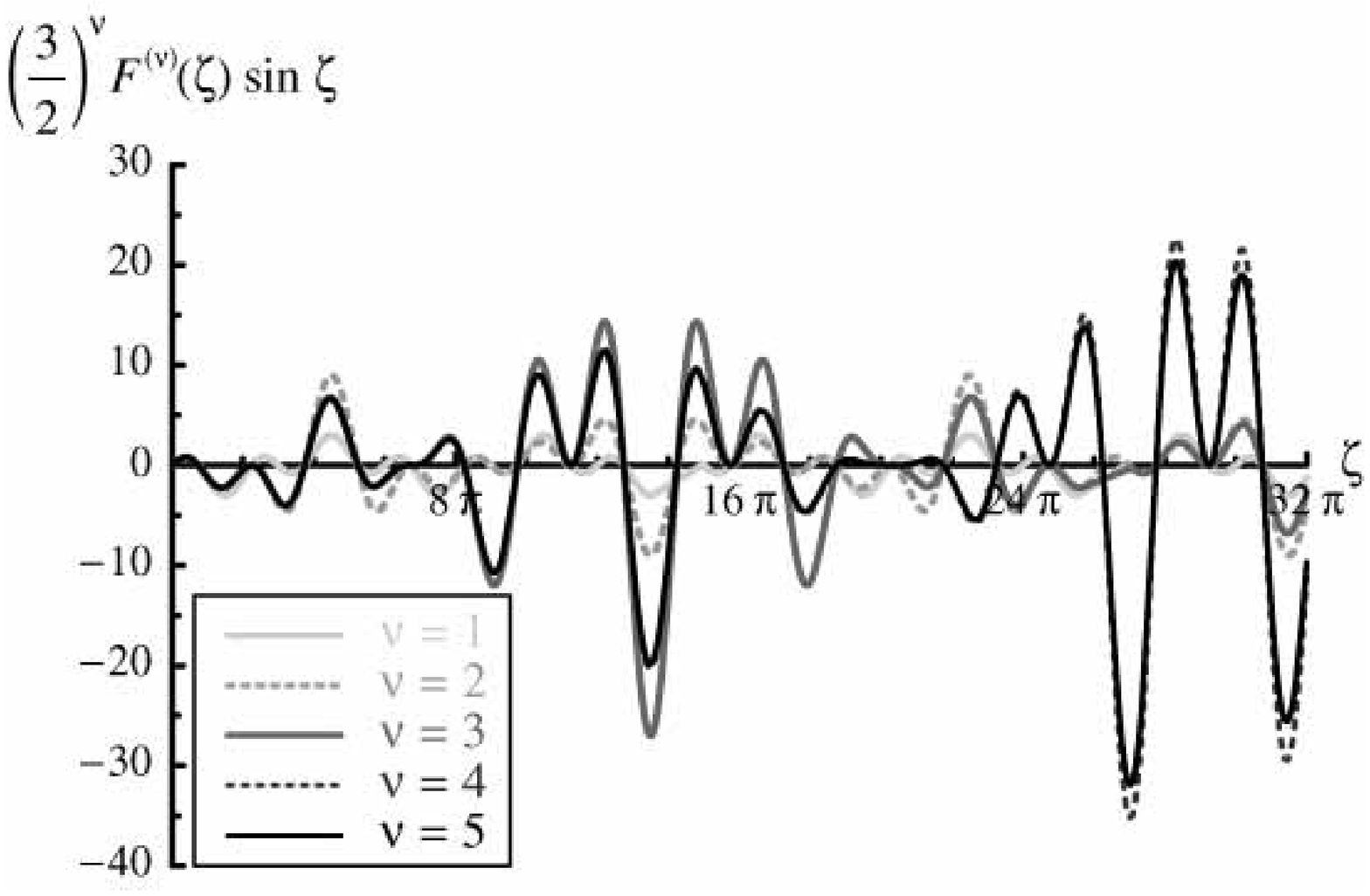}
\end{center}
\caption{We plot here $(3/2)^\nu F^{(\nu)}(\zeta)\sin\zeta$ for $1\leq\nu\leq 5$.
Curves for $\nu\geq 6$ would overlap with the curve for $\nu=5$ and be indistinguishable in this range.}
\label{Fnormalize}
\end{figure}

In the limit $\nu\to\infty$, we have $s^{(\nu)}\to 1$ and the first-order perturbation vanishes, while the coefficient $r^{(\nu)}$ depends on $\zeta$ oscillatorily.
Numerical calculation shown in Fig.~\ref{Fnormalize} suggests that the function $F^{(\nu)}(\zeta)$ has a limit of the form
\begin{equation}\label{eq3180}
f^{(\ast)}(\zeta)=\lim_{\nu\to\infty}\left(\frac{3}{2}\right)^\nu F^{(\nu)}(\zeta)\sin\zeta,
\end{equation}
although we do not know its compact expression.
For $\zeta$ at which the function $f^{(\ast)}(\zeta)/\sin\zeta$ has a moderate value, the coefficient $r^{(\nu)}$ decreases in the form
\begin{equation}\label{eq3185}
r^{(\nu)}\sim\left(\frac{2}{3}\right)^\nu\to0
\end{equation}
as $\nu\to\infty$.
The coefficient $r^{(\nu)}$, however, grows as
\begin{equation}\label{eq3190}
r^{(\nu)}(\zeta)=2^\nu\lambda
\qquad\mbox{for}\quad\zeta=3^\nu m\pi/2,
\end{equation}
where $m$ is a positive integer.
This enhancement of the reflection amplitude comes from repeated reflections at the boundaries of the Cantor set.

In order to highlight the cause of the enhancement~(\ref{eq3190}), we here compare the above result for the Cantor set with the transfer matrix for an object in which the dielectric sheets are layered periodically.
If the system of length $L$ is divided into $2l+1$ alternating sheets of the medium 1 and the medium 2, the transfer matrix is given by
\begin{eqnarray}\label{eq3200}
&&\bar{T}^{(l)}(\zeta)=\left[T_2\left(\frac{\zeta}{2l+1}\right)
T_1\left(\frac{\zeta}{2l+1}\right)\right]^l
\nonumber\\
&&\phantom{\bar{T}^{(l)}(\zeta)=T_2\left(\frac{\zeta}{2l+1}\right)}
\times T_2\left(\frac{\zeta}{2l+1}\right),
\qquad
\end{eqnarray}
where again $\zeta=k_1L$.
The Trotter formula gives its limit $l\to\infty$ in the form
\begin{equation}\label{eq3210}
\bar{T}^{(\infty)}=\E^{\I\zeta(\bar{s}\sigma_z-\I\bar{r}\sigma_x)}
\end{equation}
with
\begin{equation}\label{eq3211}
\bar{s}\equiv\frac{1+s^{(0)}}{2}
\quad\mbox{and}\quad
\bar{r}\equiv\frac{r^{(0)}}{2}.
\end{equation}
Here we have no enhancement of the reflection amplitude such as Eq.~(\ref{eq3190}).
Indeed, the enhancement~(\ref{eq3190}) occurs because the wave of $k_1=3^\nu m\pi/2L$ resonates in the cavity of the medium 1 of length $L/3$ as well as $L/3^2$, $L/3^3$ and so on (Fig.~\ref{cavity}) and the resonating waves interfere positively in every cavity.
This never happens in the periodically layered medium~(\ref{eq3200}).

We stress here that the situation causing the strong resonance is common to the Menger sponge in three spatial dimensions;
in both the Cantor set and the Menger sponge, the central part of the divided three segments is always a cavity.
We thereby suggest that the recent experiment by Takeda \textit{et al}.\cite{Takeda}\ found a strong resonance in the Menger sponge.

\section{Wave propagation within the perturbation}
\label{sec-waveprop}
\setcounter{equation}{0}

We now compute the transmission and reflection coefficients as well as seek the resonant states within the above first-order perturbation.
Let us first expand the transfer matrix $T^{(\nu)}$ with respect to $r^{(\nu)}/s^{(\nu)}$ because $s^{(\nu)}=\mathrm{O}(1)$ and $r^{(\nu)}=\mathrm{O}(\lambda)$ within the perturbation theory.
Using again the quantum analysis described in \ref{app-QA}, we have Eq.~(\ref{eq3400}), or
\begin{eqnarray}\label{eq3408}
&&T_{12}=\frac{r^{(\nu)}}{s^{(\nu)}}\sin\left(\zeta s^{(\nu)}\right),
\\
\label{eq3409}
&&T_{22}=
\E^{-\I\zeta s^{(\nu)}}
-\frac{\I\left.r^{(\nu)}\right.^2}{2\left.s^{(\nu)}\right.^2}
\left[\sin\left(\zeta s^{(\nu)}\right)
\right.
\nonumber\\
&&\phantom{T_{22}=\E^{-\I\zeta s^{(\nu)}}-}
\left.
-\zeta s^{(\nu)}\E^{-\I\zeta s^{(\nu)}}\right].
\end{eqnarray}
We thus arrive at the transmission coefficient $T$ for $\zeta\in\mathbb{R}$ in the form~(\ref{eq4020}) as
\begin{eqnarray}\label{eq3410}
T&=&\frac{1}{{T_{22}}^\ast T_{22}}
\nonumber\\
&\simeq&
\frac{1}{
1+\displaystyle{\frac{\left.r^{(\nu)}\right.^2}{\left.s^{(\nu)}\right.^2}}
\sin^2\left(\zeta s^{(\nu)}\right)
}
\nonumber\\
&\simeq&
1-\displaystyle{\frac{\left.r^{(\nu)}\right.^2}{\left.s^{(\nu)}\right.^2}}
\sin^2\left(\zeta s^{(\nu)}\right).
\end{eqnarray}
Similarly we have the reflection coefficient $R$ for $\zeta\in\mathbb{R}$  as
\begin{equation}\label{eq3420}
R=T\times \left|T_{12}\right|^2
\simeq
\frac{\left.r^{(\nu)}\right.^2}{\left.s^{(\nu)}\right.^2}\sin^2\left(\zeta s^{(\nu)}\right)
\end{equation}
Within this order, we indeed have the flux conservation $T+R=1$.
With the use of Eq.~(\ref{eq3160}), we have the expansion in $\lambda$ as
\begin{eqnarray}\label{eq3430}
R&\simeq&
\lambda^2\left.F^{(\nu)}(\zeta)\right.^2\times
\sin^2\left(\zeta + \left(\frac{2}{3}\right)^\nu\zeta\lambda \right)
\nonumber\\
&\simeq&
\lambda^2\left[F^{(\nu)}(\zeta)\sin\zeta\right]^2
\\ \label{eq3431}
&=&
2^{2\nu}\lambda^2\sin^2\frac{\zeta}{3^\nu}\prod_{k=1}^\nu\cos^2\frac{2\zeta}{3^k}.
\end{eqnarray}
and $T=1-R$.
Thus the reflection coefficient is enhanced as $R\simeq 2^{2\nu}\lambda^2$ for $\zeta=3^\nu m\pi/2$, where $m$ is a positive integer, whereas it vanishes for $\zeta$ that gives the function $f^{(\ast)}(\zeta)/\sin\zeta$ a moderate value.
\begin{figure}
\begin{center}
\includegraphics[height=0.57\textheight]{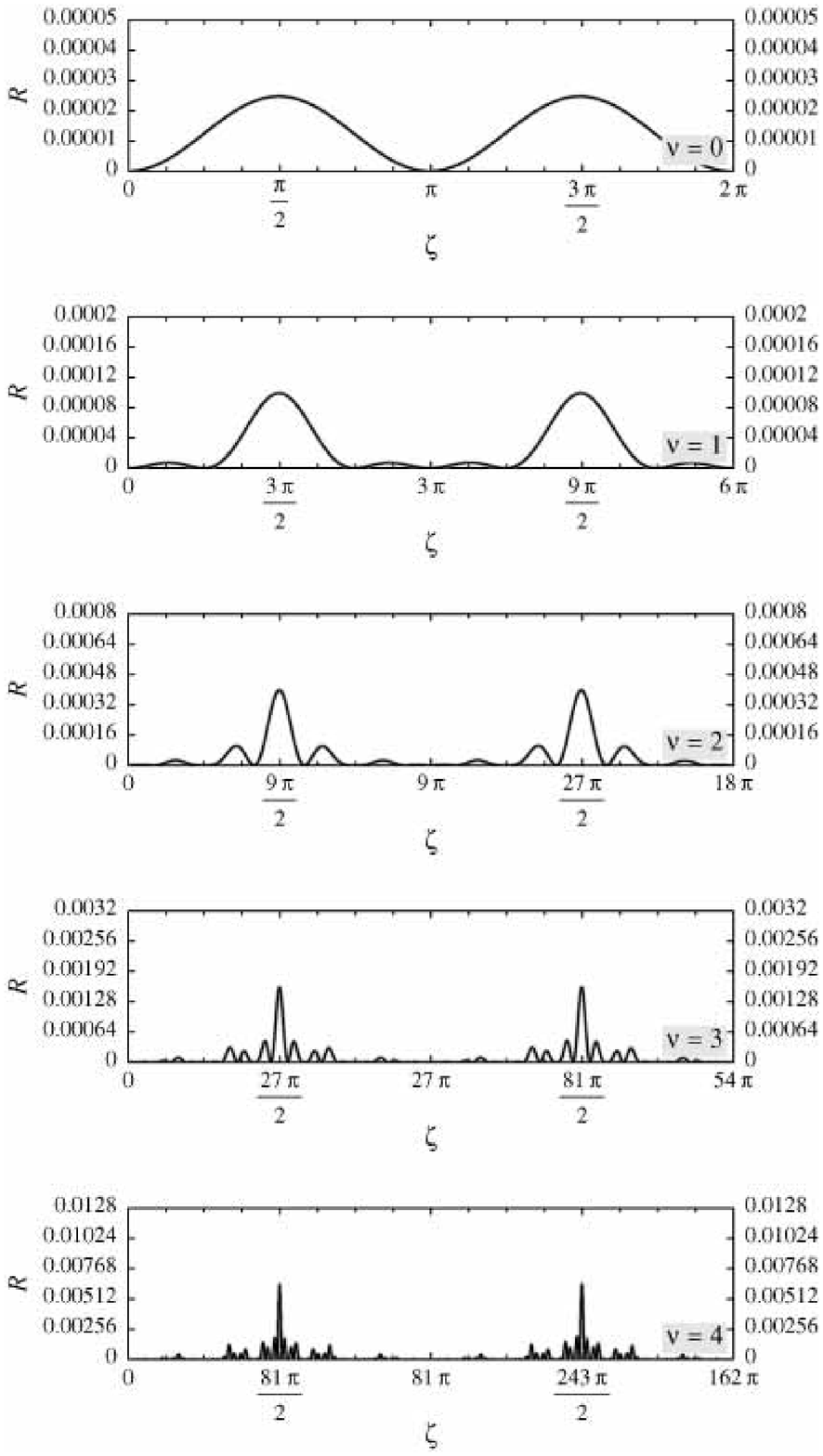}

(a)
\end{center}
\caption{The reflection coefficient $R$ due to the exact expression~(\ref{eq4020}) (the solid curves) and that due to the perturbational expression~(\ref{eq3450}) (the gray curves) are plotted for $0\leq\nu\leq4$ (from the top down to the bottom) and for (a) $\varepsilon=1.01$ and (b) $\varepsilon=1.1$.
The difference between the solid and gray curves is actually indiscernible in the panel (a).
Note that the horizontal axis of each panel is scaled by the factor $3^\nu$.}
\label{Rpert}
\end{figure}
\addtocounter{figure}{-1}
\begin{figure}
\begin{center}
\includegraphics[height=0.57\textheight]{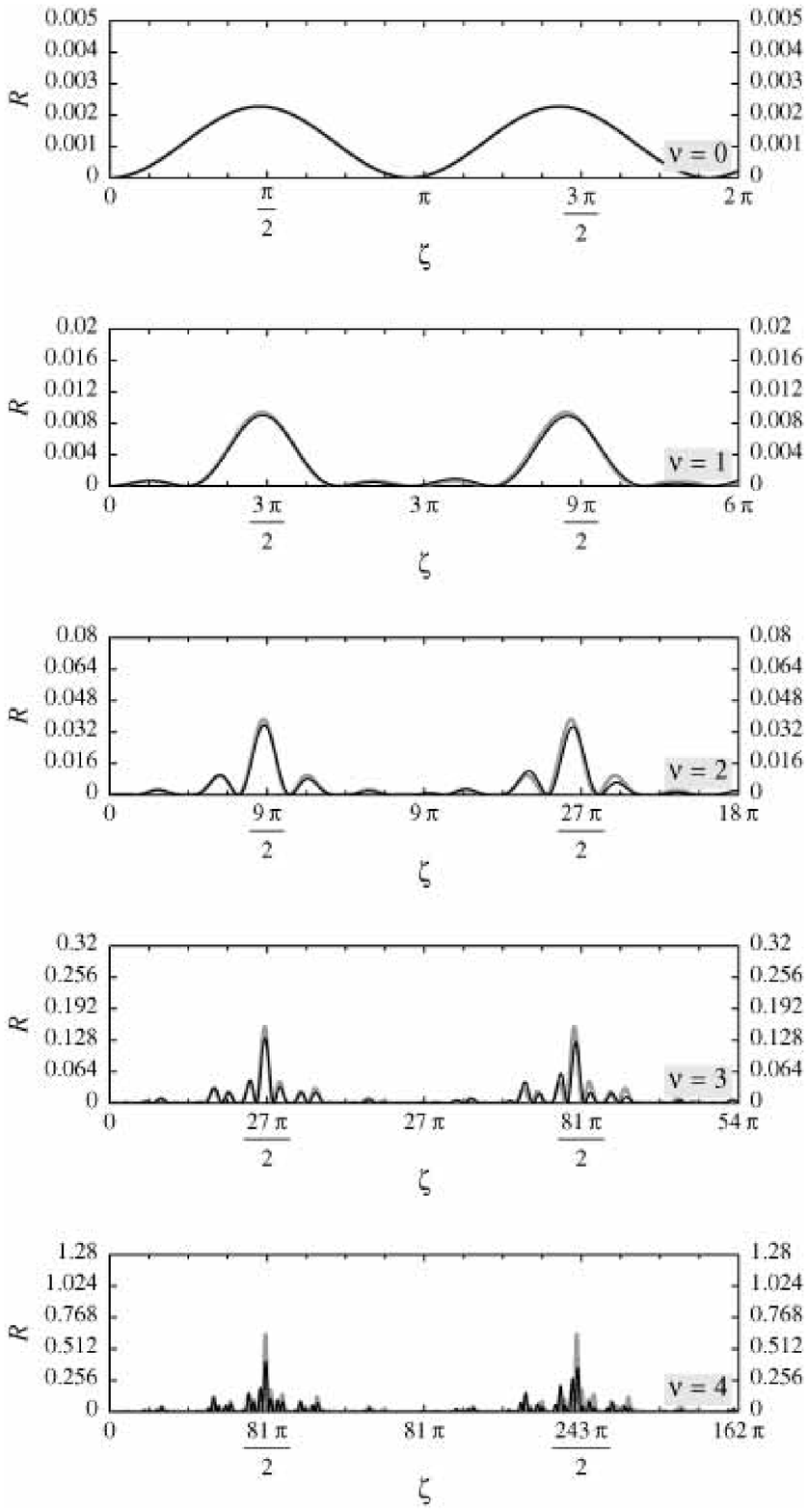}

(b)
\end{center}
\caption{\textit{Continued.}}
\end{figure}

Incidentally, the sinusoidal function in the denominator of $F^{(\nu)}(\zeta)$ in Eq.~(\ref{eq3430}) is canceled by the factor $\sin\zeta$ and hence the function $R$ does not diverge at $\zeta=m\pi$.
For the same cancellation to occur in Eq.~(\ref{eq3420}), it may be better to modify the expression to
\begin{equation}\label{eq3450}
R\simeq
\frac{\left.r^{(\nu)}\left(\zeta s^{(\nu)}\right)\right.^2}{\left.s^{(\nu)}\right.^2}\sin^2\left(\zeta s^{(\nu)}\right)
\end{equation}
so that the factor $\sin\left(\zeta s^{(\nu)}\right)$ may cancel the denominator of the function $r^{(\nu)}\left(\zeta s^{(\nu)}\right)$.
The difference between the expressions~(\ref{eq3420}) and~(\ref{eq3450}) lies only in higher orders of $\lambda$, but we found that the latter expression~(\ref{eq3450}) in fact reproduces the exact value of $R$ better, particularly when the perturbation parameter $\lambda$ gets larger.
We compared in Fig.~\ref{Rpert} the values of $R$ due to the exact expression~(\ref{eq4020}) and due to the perturbational expression~(\ref{eq3450}).
The difference is indiscernible for $\varepsilon=1.01$.
A slight difference appears at length for $\varepsilon=1.1$ and $\nu\geq2$.
\begin{figure}
\begin{center}
\includegraphics[height=0.7\textheight]{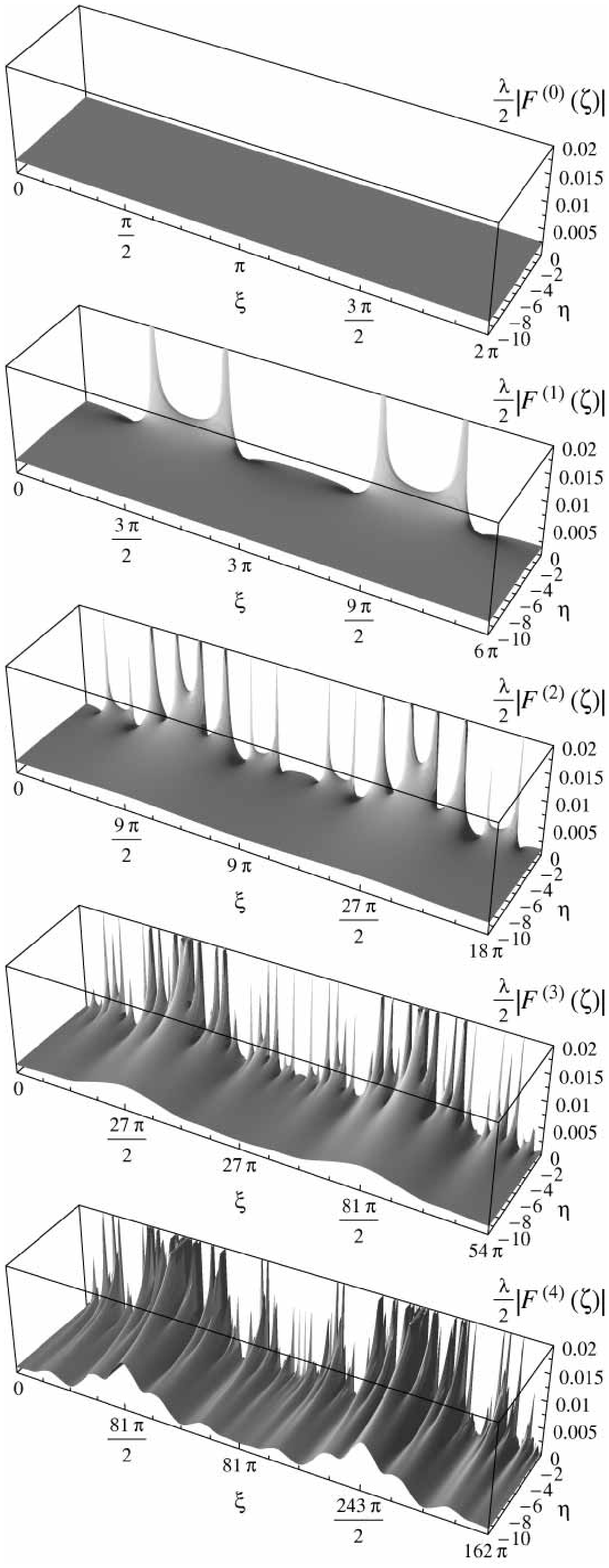}

(a)
\end{center}
\caption{(a) Three-dimensional plots of $\left|F^{(\nu)}(\zeta)\right|$ on the lower half plane of $\zeta$ for $\varepsilon=1.01$ and for $0\leq n\leq 4$ (from the top down to the bottom).
(b) Each plot of $\left|F^{(\nu)}(\zeta)\right|$ is superimposed with the function $\E^{\mathop{\mathrm{Im}}\zeta}=\E^{-|\eta|}$.}
\label{Fabs3D}
\end{figure}
\addtocounter{figure}{-1}
\begin{figure}
\begin{center}
\includegraphics[height=0.7\textheight]{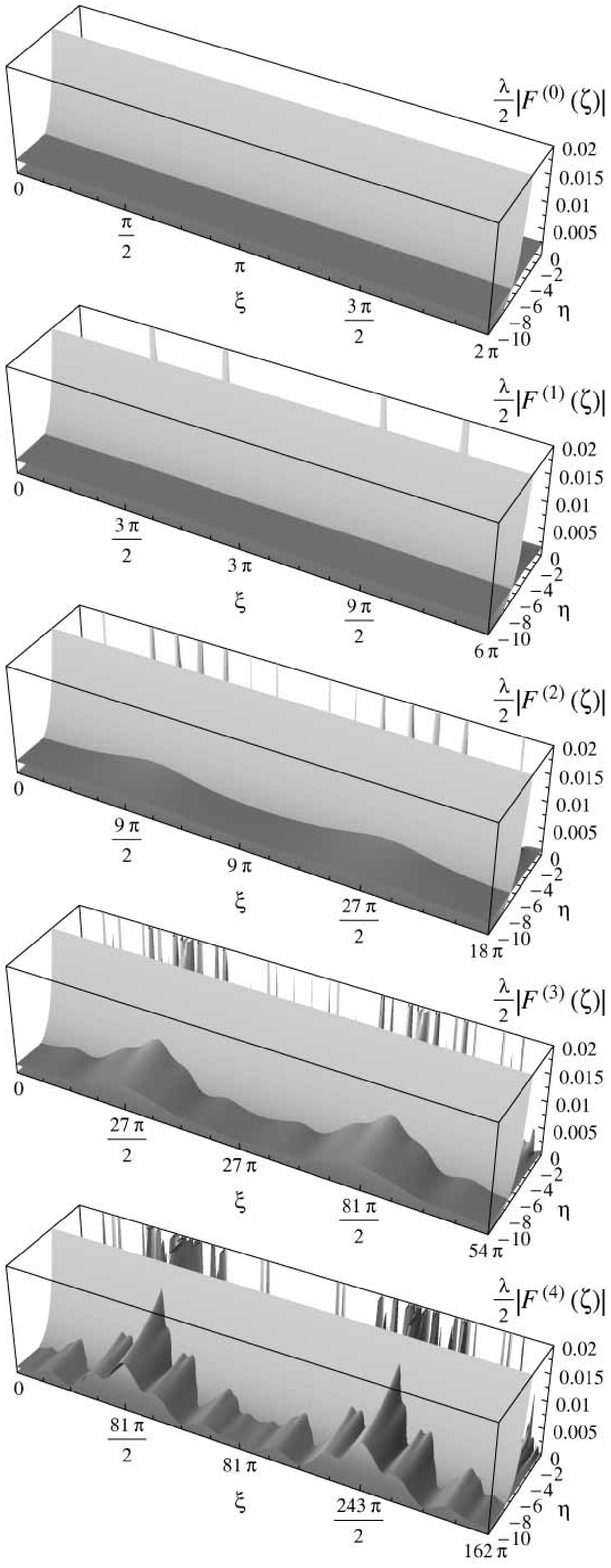}

(b)
\end{center}
\caption{\textit{Continued.}}
\end{figure}

We now look for the solution of the resonant condition~(\ref{eq4030}), or
\begin{equation}\label{eq3500}
\E^{-\I\zeta s^{(\nu)}}-\frac{\I\left.r^{(\nu)}\right.^2}{2\left.s^{(\nu)}\right.^2}
\left[\sin\left(\zeta s^{(\nu)}\right)-\zeta s^{(\nu)}\E^{-\I\zeta s^{(\nu)}}\right]=0.
\end{equation}
Rewriting the equation, we have
\begin{eqnarray}\label{eq3510}
\E^{-2\I\zeta s^{(\nu)}}
&=&\frac{\left.r^{(\nu)}\right.^2}{4\left.s^{(\nu)}\right.^2}
\left(1+\frac{\left.r^{(\nu)}\right.^2}{4\left.s^{(\nu)}\right.^2}
+\frac{\I\left.r^{(\nu)}\right.^2}{2\left.s^{(\nu)}\right.^2}
\zeta s^{(\nu)}\right)^{-1}
\nonumber\\
&\simeq&
\frac{\left.r^{(\nu)}\right.^2}{4\left.s^{(\nu)}\right.^2}.
\end{eqnarray}
We thus arrive at the equation for the resonance pole $\zeta_\mathrm{r}$ in the form
\begin{equation}\label{eq3520}
\E^{-\I\zeta_\mathrm{r} s^{(\nu)}}\simeq\pm\frac{r^{(\nu)}(\zeta_\mathrm{r})}{2s^{(\nu)}}.
\end{equation}
The use of Eq.~(\ref{eq3160}) gives the expansion in $\lambda$ as
\begin{equation}\label{eq3530}
\E^{-\I \xi_\mathrm{r}+\eta_\mathrm{r}}\simeq\pm\frac{\lambda}{2}F^{(\nu)}(\xi_\mathrm{r}+\I \eta_\mathrm{r}),
\end{equation}
where $\zeta_\mathrm{r}=\xi_\mathrm{r}+\I \eta_\mathrm{r}$.
(Note, however, that the resonance poles exist in the lower half plane: $\eta_\mathrm{r}<0$.\cite{Landau,Hatano})
We immediately have
\begin{equation}\label{eq3550}
\E^{\eta_\mathrm{r}}=\frac{\lambda}{2}\left|F^{(\nu)}(\xi_\mathrm{r}+\I \eta_\mathrm{r})\right|
\end{equation}
and
\begin{equation}\label{eq3540}
\tan \xi_\mathrm{r}=-\frac{\mathop{\mathrm{Im}}F^{(\nu)}(\xi_\mathrm{r}+\I \eta_\mathrm{r})}{\mathop{\mathrm{Re}}F^{(\nu)}(\xi_\mathrm{r}+\I \eta_\mathrm{r})}.
\end{equation}

We can understand from Eq.~(\ref{eq3550}) the fact that some of the resonance poles approach the real axis for large $\nu$.
For $\lambda=0$, the solution of Eq.~(\ref{eq3550}) is $\eta_\mathrm{r}=-\infty$.
As we turn on the perturbation parameter $\lambda$, the solution $\eta_\mathrm{r}$ takes a negatively large value.
As we stressed above, however, the function $\left|F^{(\nu)}(\zeta)\right|$ is enhanced very much for some specific values of $\zeta$ as is exemplified in Fig.~\ref{Fabs3D}~(a).
For such values of $\zeta$, the solution $\eta_\mathrm{r}$ of Eq.~(\ref{eq3550}) approaches the real axis as is shown in Fig.~\ref{Fabs3D}~(b).

Next, we show that most of the resonance poles, upon climbing up to the real axis, sway either to the right or to the left as is schematically shown in Fig.~\ref{resmove}.
Because $F^{(0)}=1$, or $\mathop{\mathrm{Im}}F^{(0)}\equiv0$, the resonance poles are aligned regularly at $\xi_\mathrm{r}=m\pi$ ($m=1,2,3,\cdots$) for $\nu=0$.
For $\nu=1$, the definition~(\ref{eq3080}) gives
\begin{equation}\label{eq3560}
F^{(1)}(\xi+\I\eta)=\frac{\E^{\I \xi-\eta}-\E^{\I \xi/3-\eta/3}+\E^{-\I \xi/3+\eta/3}-\E^{-\I \xi+\eta}}{\E^{\I \xi-\eta}-\E^{-\I \xi+\eta}},
\end{equation}
which is, as $\eta\to-\infty$, reduced to
\begin{eqnarray}\label{eq3561}
\lefteqn{
F^{(1)}(\xi+\I\eta)\simeq1-\E^{-2\I \xi/3+2\eta/3}
}
\nonumber\\
&=&\left(1-\E^{2\eta/3}\cos\frac{2\xi}{3}\right)
+\I\E^{2\eta/3}\sin\frac{2\xi}{3},
\end{eqnarray}
and hence Eq.~(\ref{eq3540}) is reduced to
\begin{equation}\label{eq3565}
\tan\xi_\mathrm{r}
\simeq
-\E^{-2\left|\eta_\mathrm{r}\right|/3}\sin\frac{2\xi_\mathrm{r}}{3}.
\end{equation}
The pole at $\xi_\mathrm{r}=\pi$ for $\nu=0$ sways to the left for $\nu=1$ because Eq.~(\ref{eq3565}) gives $\tan \xi_\mathrm{r}<0$.
Similarly, the pole at $\xi_\mathrm{r}=2\pi$ for $\nu=0$ sways to the right for $\nu=1$ because of $\tan \xi_\mathrm{r}>0$, whereas the pole at $\xi_\mathrm{r}=3\pi$ does not sway.
The same analysis gives
\begin{eqnarray}\label{eq3570}
\lefteqn{
F^{(2)}(\xi+\I\eta)=F^{(1)}(\xi+\I\eta) F^{(1)}\left(\frac{\xi+\I\eta}{3}\right)
}
\nonumber\\
&&\stackrel{\eta\to-\infty}{\simeq}
\left[1-\E^{-2\I(\xi+\I\eta)/3}\right]
\left[1-\E^{-2\I(\xi+\I\eta)/9}\right]
\nonumber\\
&&\phantom{=}
\simeq1-\E^{-2\I(\xi+\I\eta)/9}
\\ \label{eq3571}
&&\phantom{=}
=\left(1-\E^{2\eta/9}\cos\frac{2\xi}{9}\right)+\I\E^{2\eta/9}\sin\frac{2\xi}{9},
\qquad
\end{eqnarray}
or
\begin{equation}\label{eq3575}
\tan\xi_\mathrm{r}
\simeq
-\E^{-2\left|\eta_\mathrm{r}\right|/9}\sin\frac{2\xi_\mathrm{r}}{9},
\end{equation}
for $\nu=2$ and gives for an arbitrary $\nu$,
\begin{equation}\label{eq3580}
\tan\xi_\mathrm{r}
\simeq
-\E^{-2\left|\eta_\mathrm{r}\right|/3^\nu}\sin\frac{2\xi_\mathrm{r}}{3^\nu}.
\end{equation}
The exponential factor in the right-hand side of Eq.~(\ref{eq3580}) gets larger as $\nu\to\infty$.
Remember that the amplitude of the function $F^{(\nu)}(x)$ has a peak centered at $\xi=3^\nu\pi/2$.
Therefore the resonance poles to the left of the peak (where the poles climb up to the real axis) sways to the further left ($\tan\xi_\mathrm{r}<0$ for $\xi_\mathrm{r}<3^\nu\pi/2$), while the poles to the right of the peak sways to the further right ($\tan\xi_\mathrm{r}>0$ for $\xi_\mathrm{r}>3^\nu\pi/2$).
Thus we have the movement of the poles as is shown in Fig.~\ref{resmove}.

To summarize, the increase of the function $F^{(\nu)}(\zeta)$ in some areas of $\zeta$ causes the increase of the reflection coefficient $R$.
It also makes the resonance poles in the area approach the real axis, or makes the lifetimes of the resonant states longer, and at the same time repulses the resonance poles away from its peak.
Thus we end up with a wide area of a large reflection coefficient, beside which sharp peaks of the transmission coefficient appear.
We here observed the above behavior in the framework of perturbation for $\varepsilon\sim1$, but  we see the same behavior for $\varepsilon>1$ in Fig.~\ref{figTandR}~(e) and~(f).
We are going to argue below that the above behavior persists for $\varepsilon\gg1$ and for $\nu\gg1$.

\section{The strongly renormalized limit}
\label{sec-opposite}
\setcounter{equation}{0}

In the previous section, we showed in the weakly renormalized limit $\left|s^{(\nu)}\right|\gg\left|r^{(\nu)}\right|$ that the coefficient $r^{(\nu)}$ grows in the form~(\ref{eq3190}) for some values of $\zeta$ as the renormalization proceeds, and accordingly the reflection coefficient $R$ grows exponentially as is confirmed in Fig.~\ref{figTandR}~(a).
In the present section, we analyze the strongly renormalized limit $\left|s^{(\nu)}\right|\ll\left|r^{(\nu)}\right|$, and hence expand the transfer matrix with respect to $s^{(\nu)}/r^{(\nu)}$.
The computation described in \ref{app-QA} gives Eq.~(\ref{eq3630}), or
\begin{eqnarray}\label{eq3648}
T_{12}&=&
\sinh\zeta r^{(\nu)}
+\frac{\left.s^{(\nu)}\right.^2}{2\left.r^{(\nu)}\right.^2}
\sinh\zeta r^{(\nu)}
\nonumber\\
&&-\frac{\left.s^{(\nu)}\right.^2}{2\left.r^{(\nu)}\right.^2}
\zeta r^{(\nu)}\cosh\zeta r^{(\nu)}
\\ \label{eq3649}
T_{22}&=&
\cosh\zeta r^{(\nu)}
-\frac{\I s^{(\nu)}}{r^{(\nu)}}\sinh\zeta r^{(\nu)}
\nonumber\\
&&-\frac{\left.s^{(\nu)}\right.^2}{2\left.r^{(\nu)}\right.^2}
\zeta r^{(\nu)}\sinh\zeta r^{(\nu)}
\end{eqnarray}
The transmission coefficient $T$ in the present limit is therefore given by
\begin{eqnarray}\label{eq3650}
T&=&
\frac{1}{{T_{22}}^\ast T_{22}}
\nonumber\\
&\simeq&
\left[
\cosh^2\zeta r^{(\nu)}
+\frac{\left.s^{(\nu)}\right.^2}{\left.r^{(\nu)}\right.^2}\sinh^2\zeta r^{(\nu)}
\right.
\nonumber\\
&&
\left.
-\frac{\left.s^{(\nu)}\right.^2}{\left.r^{(\nu)}\right.^2}
\zeta r^{(\nu)}\sinh\zeta r^{(\nu)}\cosh\zeta r^{(\nu)}
\right]^{-1}
\nonumber\\
&\simeq&
\frac{1}{\cosh^2\zeta r^{(\nu)}}
\Biggl[1+
\nonumber\\
&&
\left.
+\frac{\left.s^{(\nu)}\right.^2}{\left.r^{(\nu)}\right.^2}\tanh\zeta r^{(\nu)}
\left(\zeta r^{(\nu)}-\tanh\zeta r^{(\nu)}\right)
\right].\qquad
\end{eqnarray}
The reflection coefficient $R$ is similarly given by
\begin{eqnarray}\label{eq3660}
R&=&T\times\left|T_{12}\right|^2
\nonumber\\
&\simeq&
T
\sinh^2\zeta r^{(\nu)}
\nonumber\\
&&\times
\left[1
+\frac{\left.s^{(\nu)}\right.^2}{\left.r^{(\nu)}\right.^2}
\left(1-\zeta r^{(\nu)}\coth\zeta r^{(\nu)}\right)
\right]
\nonumber\\
&\simeq&
\frac{1}{\cosh^2\zeta r^{(\nu)}}
\left[\sinh^2\zeta r^{(\nu)}
-\frac{\left.s^{(\nu)}\right.^2}{\left.r^{(\nu)}\right.^2}
\right.
\nonumber\\
&&\times
\tanh\zeta r^{(\nu)}
\left(\zeta r^{(\nu)}-\tanh\zeta r^{(\nu)}\right)
\Biggr].
\end{eqnarray}
Indeed we have the flux conservation $T+R=1$ in this order of perturbation.
For large $\left|r^{(\nu)}\right|$, we have $T\to0$ and $R\to1$.

The resonance condition $T_{22}=0$ now reads
\begin{equation}\label{eq3680}
\coth\zeta r^{(\nu)}=\frac{\I s^{(\nu)}}{r^{(\nu)}}+\frac{\zeta\left.s^{(\nu)}\right.^2}{\left.2r^{(\nu)}\right.^2}.
\end{equation}
By ignoring the second term of the right-hand side and rearranging further, we arrive at
\begin{equation}\label{eq3690}
\E^{2\zeta r^{(\nu)}}
=-\frac{\displaystyle 1+\frac{\I s^{(\nu)}}{r^{(\nu)}}}%
{\displaystyle 1-\frac{\I s^{(\nu)}}{r^{(\nu)}}}
\simeq -\left(1+\frac{2\I s^{(\nu)}}{r^{(\nu)}}\right),
\end{equation}
or
\begin{equation}\label{eq3700}
\E^{(\xi+\I\eta) r^{(\nu)}}\simeq\I\left(1+\frac{\I s^{(\nu)}}{r^{(\nu)}}\right),
\end{equation}
where we put $\zeta=\xi+\I\eta$.
We hence have
\begin{equation}\label{eq3710}
\E^{2\xi r^{(\nu)}}\simeq1,
\qquad\mbox{or}\quad
r^{(\nu)}\simeq0,
\end{equation}
which appears to be inconsistent with the first assumption $\left|s^{(\nu)}\right|\ll\left|r^{(\nu)}\right|$.
This means that the resonance poles do not exist in this limit.

Thus we again observe the following behavior:
in the area where $r^{(\nu)}$ grows, the poles move toward the real axis, but at the same time avoid the area of large $\left|r^{(\nu)}\right|$, opening up a wide gap free of resonance poles.
The reflection coefficient $R$ is close to unity in the wide gap, while the resonance poles just beside the gap give very sharp peaks of the transmission coefficient $T$.

\section{Trivial fixed point}
\label{sec-trivial}
\setcounter{equation}{0}

We showed in \S\ref{sec-pert} that the reflection amplitude $r^{(\nu)}(\zeta)$ grows exponentially for some $\zeta$ as in Eq.~(\ref{eq3190}), but decays exponentially for other $\zeta$ as in Eq.~(\ref{eq3185}).
In the previous section, we analyze the limit of the growing reflection amplitude.
We examine in the present section, on the other hand, the trivial fixed point
\begin{equation}\label{eq3790}
r^{(\nu)}(\zeta)\sim\left(\frac{2}{3}\right)^\nu\to 0.
\end{equation}
We showed in \S\S\ref{sec-waveprop} and~\ref{sec-opposite} that some of the resonance poles climb up to the real axis of the wave number near a peak of the reflection amplitude $r^{(\nu)}$ but are eventually repulsed away from the peak.
We argue here that such resonance poles in fact approaches the trivial fixed point~(\ref{eq3790}).

\begin{figure}
\begin{center}
\includegraphics[height=0.7\textheight]{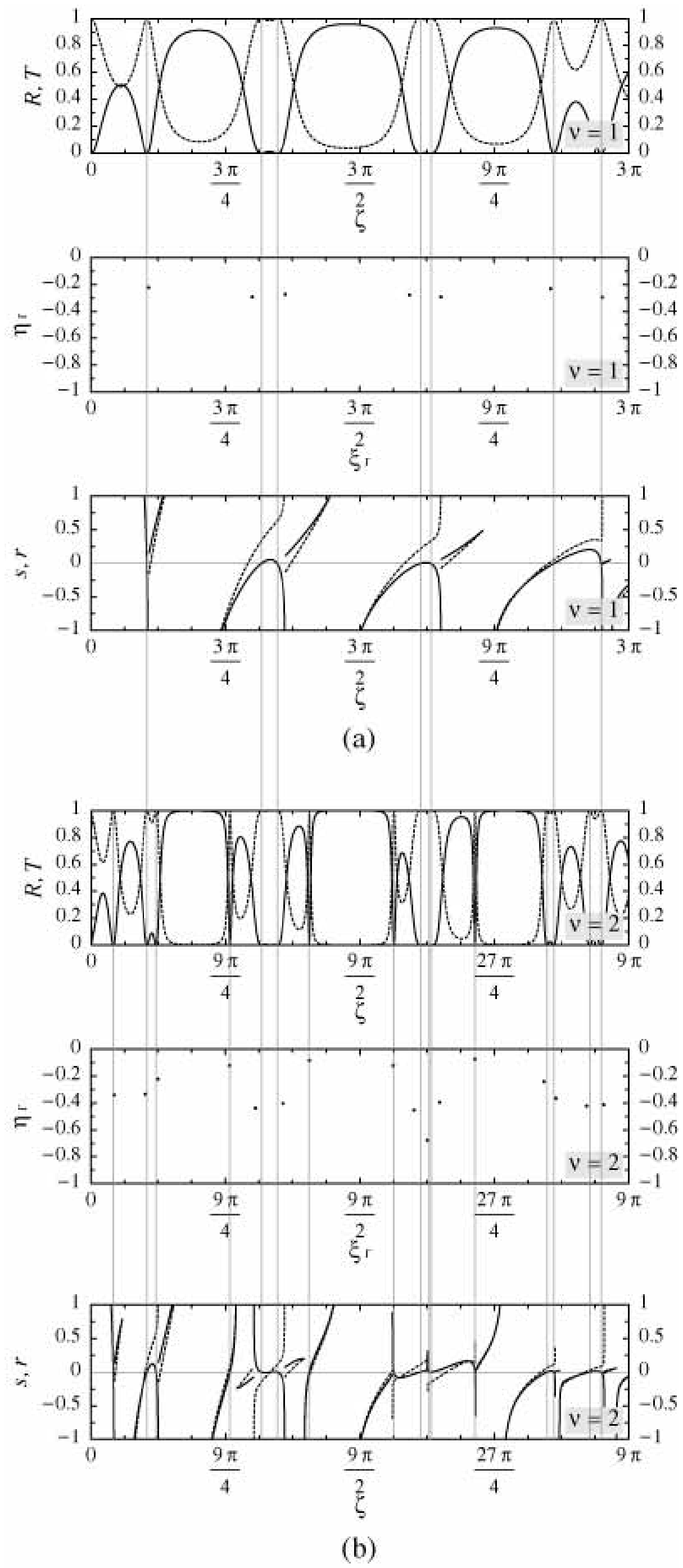}
\end{center}
\caption{The bottom graph of each of the panel (a) $\nu=1$ and the panel (b) $\nu=2$ gives the $\zeta$ dependence of $s^{(\nu)}$ (the broken curves) and $r^{(\nu)}$ (the solid curves) for $\varepsilon=10$.
The mid graph of each panel shows the locations of the resonance poles and the top graph of each panel shows the $\zeta$ dependence of the transmission and reflection coefficients.
The mid and the top graphs are the same as the ones in Fig.~\ref{figTandR}~(e) and~(f).
The vertical gray lines indicate the zeros of $r^{(\nu)}$.
Some of the zeros are common to the both panels.}
\label{figvar}
\end{figure}

The transfer matrix becomes diagonal at the point where $r^{(\nu)}(\zeta)=0$:
\begin{equation}\label{eq3800}
T^{(\nu)}(\zeta)
=\E^{\I\zeta s^{(\nu)}\sigma_z}
=\left(\begin{array}{cc}
\E^{\I\zeta s^{(\nu)}} & \\
& \E^{-\I\zeta s^{(\nu)}}
\end{array}\right),
\end{equation}
which is immediately followed by
\begin{equation}\label{eq3805}
T=1
\quad\mbox{and}\quad
R=0.
\end{equation}
Thus the transmission coefficient has a sharp peak at the real solutions of the equation $r^{(\nu)}(\zeta)=0$ as is demonstrated in Fig.~\ref{figvar}.
We also know that transmission coefficient has a sharp peak in the case that a resonance pole is close to the real axis.
Combining the above two observations, we conclude that the resonance pole,
\begin{enumerate}
\item first climbing up a peak of $r^{(\nu)}$ toward the real axis,
\item next sliding off the peak sideways,
\item eventually approaches a solution of the equation $r^{(\nu)}(\zeta)=0$ on the real axis.
\end{enumerate}
We indeed observe in Fig.~\ref{figvar} that the resonance poles right next to the wide regions of $R\sim1$ are particularly close to the real axis and are accompanied by a very narrow peak of $T$.
The poles away from the regions $R\sim1$, on the other hand, are not close to peaks of $r^{(\nu)}$, and hence are not close to the fixed point $r^{(\nu)}=0$ on the real axis.
(We have $T\to\infty$ at the resonance pole, while we have $T=1$ on the real axis,
which appears to lead to a contradiction when the resonance pole approaches the real axis.
They are reconciled if $\D s^{(\nu)}(\zeta)/\D\zeta$ diverges at the pole.)

Incidentally, the recursion formula of the transfer matrix shows that the matrix $T^{(\nu+1)}(\zeta)$ is diagonal if the matrix $T^{(\nu)}(\zeta/3)$ is diagonal.
This means $r^{(\nu+1)}(\zeta)=0$ if $r^{(\nu)}(\zeta/3)=0$.
This is also demonstrated in Fig.~\ref{figvar};
some of the solutions $r^{(2)}(\zeta)=0$ are indeed given by $r^{(1)}(\zeta/3)=0$.

\section{Discussions}
\label{sec-discuss}
\setcounter{equation}{0}

To summarize, we have analyzed how the resonance poles $\zeta_\mathrm{r}$ move in the complex wave-number plane as we progress to higher generations of the Cantor set.
We conclude that the fractal structure of the medium produces resonance poles very close to the real axis of the wave number with a wide region of the wave number without resonant states.
As a result, we have sharp peaks of the transmission coefficient ($T\sim1$) and a wide region of nearly total reflection ($R\sim1$) between them.

Although we carried out our analysis only in one spatial dimension, we expect that the above conclusion itself is independent of the dimensionality.
As we emphasized in Fig.~\ref{cavity} and below Eq.~(\ref{eq3190}), the strong resonance is essentially due to the fractal structure in which the central part of the divided three segments is always a cavity.
This structure is common to the Menger sponge in three spatial dimensions.
We thereby suggest that the recent experiment by Takeda \textit{et al}.\cite{Takeda}\ found a strong resonance in the Menger sponge.

It is then quite natural for them to observe dips both in the transmission amplitude and the reflection amplitude.
In the resonant state, the incident wave is scattered strongly and repeatedly inside the cavity for the duration of the resonant lifetime.
Therefore the outgoing wave can have an amplitude over quite a wide range of the scattering angle in three dimensions.
We remind the readers of the fact that, in one spatial dimension, the resonant wave function itself is either symmetric or anti-symmetric and the field strengths on the left and the right are therefore equal.
Close to the resonance pole, the field distribution is almost symmetric at the resonant frequency on the real axis.
Indeed, the resonant scattering observed in Fig.~\ref{fieldres} is almost symmetric and independent of the direction of the incident wave.
We stress here that this happens for the three-dimensional resonance as well.
For the Menger sponge embedded in three spatial dimensions, some of the resonant states have outgoing waves of the same amplitude in all six directions.
At the resonant frequency close to such resonant states but on the real axis, the scattered wave has a strong amplitude even in the directions perpendicular to the incident wave vector.
We hence claim that the lost fraction of the transmission and reflection amplitudes in the experiment by Takeda \textit{et al}.\cite{Takeda}\ must have been emitted in the directions normal to the incident direction.

\section*{Acknowledgments}
The author is grateful to Prof. K. Ohtaka for encouragement, discussions and information on photonic crystals.
Financial support from the Sumitomo Foundation is gratefully acknowledged.
The study is also supported partially by Grant-in-Aid for Scientific Research (No.17340115) from the Ministry of Education, Culture, Sports, Science and Technology.

\renewcommand{\thesection}{Appendix \Alph{section}}
\renewcommand{\theequation}{\Alph{section}.\arabic{equation}}
\setcounter{section}{0}
\section{Suzuki's quantum analysis}
\label{app-QA}
\setcounter{equation}{0}

In the present Appendix, we give details of the calculations producing Eqs.~(\ref{eq3061}),~(\ref{eq3140}),~(\ref{eq3408})--(\ref{eq3409}) and~(\ref{eq3648})--(\ref{eq3649}).
We heavily used M.~Suzuki's ``quantum analysis"\cite{Suzuki1,Suzuki2} in the calculation.
See Hatano and Suzuki\cite{India} for a recent review.

The main feature of the quantum analysis is the differentiation of an operator functional $f(A)$ by an operator $A$.
The derivative $\D f(A)/\D A$ is given as a hyperoperator mapping $\D A$ to $\D f(A)$, defined in
\begin{equation}\label{eq9000}
\D f(A)=\lim_{h\to0}\frac{f(A+h\D A)-f(A)}{h}
\equiv
\frac{\D f(A)}{\D A}\cdot \D A,
\end{equation}
where the convergence of the limit operation is in the sense of the norm convergence uniform with respect to the arbitrary operator $\D A$.
It should be noted that the hyperoperator $\D f(A)/\D A$ is expressed in terms of the operator $A$ itself and its inner derivation $\delta_A$.
The inner derivation is a hyperoperator giving the operation of nothing other than the commutation relation:
\begin{equation}\label{eq9010}
\delta_A B \equiv \left[A,B\right].
\end{equation}

Once we obtain the operator derivative~(\ref{eq9000}), we can calculate the parameter derivative of the operator functional $f(A(x))$ in the form
\begin{equation}\label{eq9020}
\frac{\D f(A(x))}{\D x}=\frac{\D f(A)}{\D A}\cdot
\frac{\D A(x)}{\D x}.
\end{equation}
A particularly important formula given by the quantum analysis is the operator derivative of the exponential of an operator:\cite{Suzuki2}
\begin{equation}\label{eq9030}
\frac{\D \E^A}{\D A}=\E^A \frac{1-\E^{-\delta_A}}{\delta_A}
\equiv
\E^A\sum_{n=1}^\infty\frac{(-1)^{n-1}}{n!}\left.\delta_A\right.^{n-1}.
\end{equation}
We use this formula repeatedly hereafter.

Let us differentiate the both sides of Eq.~(\ref{eq2050}) with respect to $\lambda$.
The formula~(\ref{eq9030}) with Eq.~(\ref{eq9020}) gives
\begin{eqnarray}\label{eq2060}
\lefteqn{
\frac{\D}{\D\lambda}\E^{\I\zeta(A+\lambda C^{(0)})/3}
}
\nonumber\\
&=&\frac{\I\zeta}{3}\E^{\I\zeta(A+\lambda C^{(0)})/3}
\nonumber\\
&&\times
\frac{1-\exp\left[-\delta_{\I\zeta(A+\lambda C^{(0)})/3}\right]}%
{\delta_{\I\zeta(A+\lambda C^{(0)})/3}}
C^{(0)}.
\end{eqnarray}
Note that $\delta_{xA}=x\delta_A$.
Thus we have
\begin{eqnarray}\label{eq2080}
\lefteqn{
\left.
\frac{\D}{\D\lambda}\E^{\I\zeta(A+\lambda C^{(0)})/3}
\right|_{\lambda=0}
=\frac{\I\zeta}{3}\E^{\I\zeta A/3}\frac{1-\E^{-(\I\zeta/3)\delta_A}}{(\I\zeta/3)\delta_A}C^{(0)}
}
\nonumber\\
&=&-\E^{\I\zeta A/3}\frac{\E^{-(\I\zeta/3)\delta_A}-1}{\delta_A}C^{(0)}
\\
\label{eq2085}
&=&-\E^{\I\zeta A/3}\frac{\E^{-(\I\zeta/3)\delta_A}-1}{\delta_A}C^{(0)}\E^{-\I\zeta A/3}\E^{\I\zeta A/3}
\nonumber\\
&=&-\E^{(\I\zeta/3)\delta_A}\frac{\E^{-(\I\zeta/3)\delta_A}-1}{\delta_A}C^{(0)}\E^{\I\zeta A/3}
\nonumber\\
&=&-\frac{1-\E^{(\I\zeta/3)\delta_A}}{\delta_A}C^{(0)}\E^{\I\zeta A/3}.
\end{eqnarray}
Here we used the identity
\begin{equation}\label{eq3005}
\E^A B \E^{-A} = \E^{\delta_A}B
\end{equation}
in moving from the third line to the fourth line.
Therefore, the differentiation of the left-hand side of Eq.~(\ref{eq2050}) yields
\begin{eqnarray}\label{eq3010}
\lefteqn{
\left.
\frac{\D}{\D\lambda}
\E^{\I\zeta(A+\lambda C^{(0)})/3}\E^{\I\zeta A/3}\E^{\I\zeta(A+\lambda C^{(0)})/3}
\right|_{\lambda=0}
}
\nonumber\\
&&\qquad
=
-\frac{1-\E^{(\I\zeta/3)\delta_A}}{\delta_A}C^{(0)}\E^{\I\zeta A}
\nonumber\\
&&\qquad\phantom{=}
-\E^{\I\zeta A}\frac{\E^{-(\I\zeta/3)\delta_A}-1}{\delta_A}C^{(0)},
\end{eqnarray}
where we used both of the expressions~(\ref{eq2080}) and~(\ref{eq2085}).
On the other hand, the differentiation of the right-hand side of Eq.~(\ref{eq2050}) yields
\begin{eqnarray}\label{eq3020}
\left.
\frac{\D}{\D\lambda}
\E^{\I\zeta (A+\lambda C^{(1)} + \mathrm{O}(\lambda^2))}
\right|_{\lambda=0}
=-\E^{\I\zeta A}\frac{\E^{-\I\zeta\delta_A}-1}{\delta_A}C^{(1)}
\end{eqnarray}
because of the formula~(\ref{eq9030}).
Comparison of Eq.~(\ref{eq3010}) with Eq.~(\ref{eq3020}) is followed by
\begin{eqnarray}\label{eq3030}
\lefteqn{
C^{(1)}=\frac{\delta_A}{\E^{-\I\zeta\delta_A}-1}\E^{-\I\zeta A}
}
\nonumber\\
&&
\times
\left(
\frac{1-\E^{(\I\zeta/3)\delta_A}}{\delta_A}C^{(0)}\E^{\I\zeta A}
\right.
\nonumber\\
&&
\phantom{\frac{\delta_A}{\E^{-\I\zeta\delta_A}-1}\E^{-\I\zeta A}}
\left.
+\E^{\I\zeta A}\frac{\E^{-(\I\zeta/3)\delta_A}-1}{\delta_A}C^{(0)}
\right)
\nonumber\\
&&=
\frac{1}{\E^{-\I\zeta\delta_A}-1}
\left[\E^{-\I\zeta \delta_A}\left(1-\E^{(\I\zeta/3)\delta_A}\right)C^{(0)}
\right.
\nonumber\\
&&\phantom{\frac{1}{\E^{-\I\zeta\delta_A}-1}}
\left.
+\left(\E^{-(\I\zeta/3)\delta_A}-1\right)C^{(0)}
\right]
\nonumber\\
&&=
\frac{\E^{-\I\zeta \delta_A}-\E^{-(2\I\zeta/3)\delta_A}
+\E^{-(\I\zeta/3)\delta_A}-1}{\E^{-\I\zeta\delta_A}-1}
C^{(0)},
\qquad\quad
\end{eqnarray}
where we again used Eq.~(\ref{eq3005}) in moving from the first line to the second line.
The final expression is an even function of $\zeta\delta_A$.
Therefore the Taylor expansion with respect to $\zeta\delta_A$ gives only even-order terms:
\begin{equation}\label{eq3040}
C^{(1)}=\sum_{k: \mathrm{even}}c_k\left(-\I\zeta\right)^k{\delta_A}^kC^{(0)},
\end{equation}
where the coefficients $c_k$ are given by the Taylor expansion
\begin{equation}\label{eq3050}
\frac{\E^x-\E^{2x/3}+\E^{x/3}-1}{\E^x-1}
=\sum_{k=0}^\infty c_kx^k
\end{equation}
with $c_0=2/3$.

Using the definition~(\ref{eq2040}) explicitly here, we have
\begin{eqnarray}\label{eq2090}
\delta_A C^{(0)}&=&\left[\sigma_z,\sigma_z-\I\sigma_x\right]
\nonumber\\
&=&-\I\left[\sigma_z,\sigma_x\right]=2\sigma_y,
\\
\label{eq2100}
{\delta_A}^2C^{(0)}&=&\left[\sigma_z,2\sigma_y\right]=-4\I\sigma_x,
\\
\label{eq2110}
{\delta_A}^3C^{(0)}&=&\left[\sigma_z,-4\I\sigma_x\right]=8\sigma_y,\cdots
\end{eqnarray}
or
\begin{equation}\label{eq2120}
{\delta_A}^kC^{(0)}=
\left\{
\begin{array}{ll}
2^k\sigma_y & \mbox{for odd $k$},\\
\delta_{k0}\sigma_z-2^k\I\sigma_x & \mbox{for even $k$}.
\end{array}
\right.
\end{equation}
Substituting the result~(\ref{eq2120}) for the terms in the summation of Eq.~(\ref{eq3040}), we arrive at
\begin{eqnarray}\label{eq3060}
C^{(1)}&=&\frac{2}{3}\sigma_z-\I\sigma_x\sum_{k: \mathrm{even}}c_k\left(-2\I\zeta\right)^k
\nonumber\\
&=&\frac{2}{3}\sigma_z-
\frac{\E^{-2\I\zeta}-\E^{-4\I\zeta/3}+\E^{-2\I\zeta/3}-1}{\E^{-2\I\zeta}-1}
\I\sigma_x
\nonumber\\
&=&
\frac{2}{3}\sigma_z-\I F^{(1)}(\zeta)\sigma_x,
\end{eqnarray}
or Eq.~(\ref{eq3061}), where $F^{(1)}(\zeta)$ is defined by Eq.~(\ref{eq3081}).

The next renormalization transformation is obtained by replacing $C^{(0)}$ with $C^{(1)}$ and $C^{(1)}$ with $C^{(2)}$ in Eq.~(\ref{eq3030}).
In the replaced $C^{(1)}$, the argument of the function $F^{(1)}$ is $\zeta/3$.
Instead of Eqs.~(\ref{eq2090})--(\ref{eq2120}), we now have
\begin{eqnarray}\label{eq3090}
\delta_A C^{(1)}&=&\left[\sigma_z,\frac{2}{3}\sigma_z-\I F^{(1)}\left(\frac{\zeta}{3}\right)\sigma_x\right]
\nonumber\\
&=&-\I F^{(1)}\left(\frac{\zeta}{3}\right)\left[\sigma_z,\sigma_x\right]
\nonumber\\
&=&2F^{(1)}\left(\frac{\zeta}{3}\right)\sigma_y,
\\
\label{eq3100}
{\delta_A}^2C^{(1)}&=&\left[\sigma_z,2F^{(1)}\left(\frac{\zeta}{3}\right)\sigma_y\right]
\nonumber\\
&=&-4\I F^{(1)}\left(\frac{\zeta}{3}\right)\sigma_x,
\\
\label{eq3110}
{\delta_A}^3C^{(1)}&=&\left[\sigma_z,-4\I F^{(1)}\left(\frac{\zeta}{3}\right)\sigma_x\right]
\nonumber\\
&=&8F^{(1)}\left(\frac{\zeta}{3}\right)\sigma_y,\cdots
\end{eqnarray}
or
\begin{equation}\label{eq3120}
{\delta_A}^kC^{(1)}=
\left\{
\begin{array}{ll}
{\displaystyle 2^kF^{(1)}\left(\frac{\zeta}{3}\right)\sigma_y} & \mbox{for odd $k$},\\
{\displaystyle \frac{2}{3}\delta_{k0}\sigma_z-2^k\I F^{(1)}\left(\frac{\zeta}{3}\right)\sigma_x} & \mbox{for even $k$}.
\end{array}
\right.
\end{equation}
We thus have
\begin{eqnarray}\label{eq3130}
C^{(2)}&=&\left(\frac{2}{3}\right)^2\sigma_z
-\I F^{(1)}(\zeta)F^{(1)}\left(\frac{\zeta}{3}\right)\sigma_x
\nonumber\\
&=&\left(\frac{2}{3}\right)^2\sigma_z
-\I F^{(2)}(\zeta)\sigma_x,
\end{eqnarray}
or Eq.~(\ref{eq3140}), where $F^{(2)}(\zeta)$ is defined by (\ref{eq3150}).

\onecolumn
We now expand the transfer matrix with respect to $r^{(\nu)}/s^{(\nu)}$.
Calculation similar to Eqs.~(\ref{eq2085}) yields 
\begin{eqnarray}\label{eq3300}
\frac{\D}{\D r^{(\nu)}}T^{(\nu)}
&=&
\E^{\I\zeta(s^{(\nu)}\sigma_z-\I r^{(\nu)}\sigma_x)}
\frac{1-\exp\left[-\I\zeta\left(s^{(\nu)}\delta_{\sigma_z}-\I r^{(\nu)}\delta_{\sigma_x}\right)\right]}{\I\zeta\left(s^{(\nu)}\delta_{\sigma_z}-\I r^{(\nu)}\delta_{\sigma_x}\right)}
\zeta\sigma_x
\nonumber\\
&=&
\zeta\E^{\I\zeta(s^{(\nu)}\sigma_z-\I r^{(\nu)}\sigma_x)}
\sum_{k=1}^\infty
\frac{(-\I\zeta)^{k-1}}{k!}
\left(s^{(\nu)}\delta_{\sigma_z}-\I r^{(\nu)}\delta_{\sigma_x}\right)^{k-1}\sigma_x.
\end{eqnarray}
The zeroth-order term of $r^{(\nu)}$ in the summation on the right-hand side is
\begin{equation}\label{eq3310}
\sum_{k=1}^\infty
\frac{(-\I\zeta s^{(\nu)})^{k-1}}{k!}
{\delta_{\sigma_z}}^{k-1}\sigma_x.
\end{equation}
The same computation as in Eqs.~(\ref{eq2090})--(\ref{eq2110}) gives
\begin{equation}\label{eq3320}
{\delta_{\sigma_z}}^k\sigma_x=\left\{
\begin{array}{ll}
2^k\I\sigma_y & \qquad\mbox{for odd $k$},\\
2^k\sigma_x & \qquad\mbox{for even $k$},
\end{array}
\right.
\end{equation}
and hence we have
\begin{equation}\label{eq3330}
\mbox{Eq.~(\ref{eq3310})}
=
\sum_{k=1,3,5,\cdots}^\infty
\frac{(-2\I\zeta s^{(\nu)})^{k-1}}{k!}\sigma_x
+\sum_{k=2,4,6,\cdots}^\infty
\frac{(-2\I\zeta s^{(\nu)})^{k-1}}{k!}\I\sigma_y
=
\frac{1}{2\zeta s^{(\nu)}}\left\{
\sigma_x\sin2\zeta s^{(\nu)}
+\sigma_y\left[1-\cos2\zeta s^{(\nu)}\right]
\right\}.
\end{equation}
The first-order term of $r^{(\nu)}$ in the summation on the right-hand side of Eq.~(\ref{eq3300}) is
\begin{equation}\label{eq3340}
-\I\sum_{k=2}^\infty
\frac{(-\I\zeta )^{k-1}}{k!}\left.s^{(\nu)}\right.^{k-2}r^{(\nu)}
\sum_{l=0}^{k-2}
{\delta_{\sigma_z}}^{k-l-2}\delta_{\sigma_x}{\delta_{\sigma_z}}^l\sigma_x.
\end{equation}
Because of Eq.~(\ref{eq3320}), we have
\begin{equation}\label{eq3350}
\delta_{\sigma_x}{\delta_{\sigma_z}}^l\sigma_x=\left\{
\begin{array}{ll}
-2^{l+1}\sigma_z & \qquad\mbox{for odd $l$},\\
0 & \qquad\mbox{for even $l$}.
\end{array}\right.
\end{equation}
Therefore, the only remaining terms in the summations of Eq.~(\ref{eq3340}) are for $l=k-2=$ odd with the result $\delta_{\sigma_x}{\delta_{\sigma_z}}^{k-2}\sigma_x=-2^{k-1}\sigma_z$.
Thus we have
\begin{equation}\label{eq3360}
\mbox{Eq.~(\ref{eq3340})}
=
\I r^{(\nu)}\sum_{k=3,5,7,\cdots}^\infty
\frac{(-\I\zeta )^{k-1}}{k!}\left.s^{(\nu)}\right.^{k-2}\times2^{k-1}\sigma_z
=\frac{r^{(\nu)}}{s^{(\nu)}}
\left[
\frac{\sin\left(2\zeta s^{(\nu)}\right)}{2\zeta s^{(\nu)}}-1
\right]\I\sigma_z.
\end{equation}
To summarize, we have
\begin{equation}\label{eq3370}
\frac{\D}{\D r^{(\nu)}}T^{(\nu)}
=
\frac{\E^{\I\zeta(s^{(\nu)}\sigma_z-\I r^{(\nu)}\sigma_x)}}{2s^{(\nu)}}
\left\{
\sigma_x\sin2\zeta s^{(\nu)}
+\sigma_y\left[1-\cos2\zeta s^{(\nu)}\right]
+\I\sigma_z\frac{r^{(\nu)}}{s^{(\nu)}}
\left[
\sin\left(2\zeta s^{(\nu)}\right)-2\zeta s^{(\nu)}
\right]
\right\}
+\mathrm{O}\left(\left.r^{(\nu)}\right.^2\right),
\end{equation}
and then have
\begin{eqnarray}\label{eq3380}
\frac{\D^2}{\left.\D r^{(\nu)}\right.^2}T^{(\nu)}
&=&
\frac{\E^{\I\zeta(s^{(\nu)}\sigma_z-\I r^{(\nu)}\sigma_x)}}{4\left.s^{(\nu)}\right.^2}
\left\{
\sigma_x\sin2\zeta s^{(\nu)}
+\sigma_y\left[1-\cos2\zeta s^{(\nu)}\right]
+\I\sigma_z\frac{r^{(\nu)}}{s^{(\nu)}}
\left[
\sin\left(2\zeta s^{(\nu)}\right)-2\zeta s^{(\nu)}
\right]
\right\}^2
\nonumber\\
&&
+\frac{\E^{\I\zeta(s^{(\nu)}\sigma_z-\I r^{(\nu)}\sigma_x)}}{2\left.s^{(\nu)}\right.^2}
\left[
\sin\left(2\zeta s^{(\nu)}\right)-2\zeta s^{(\nu)}
\right]\I\sigma_z
+\mathrm{O}\left(r^{(\nu)}\right)
\end{eqnarray}
We finally obtain
\begin{eqnarray}\label{eq3390}
\left.\frac{\D}{\D r^{(\nu)}}T^{(\nu)}\right|_{r^{(\nu)}=0}
&=&
\frac{\E^{\I\zeta s^{(\nu)}\sigma_z}}{2s^{(\nu)}}
\left\{
\sigma_x\sin2\zeta s^{(\nu)}
+\sigma_y\left[1-\cos2\zeta s^{(\nu)}\right]
\right\},
\\
\left.\frac{\D^2}{\left.\D r^{(\nu)}\right.^2}T^{(\nu)}\right|_{r^{(\nu)}=0}
&=&
\frac{\E^{\I\zeta s^{(\nu)}\sigma_z}}{4\left.s^{(\nu)}\right.^2}\left\{
\sigma_x\sin2\zeta s^{(\nu)}
+\sigma_y\left[1-\cos2\zeta s^{(\nu)}\right]
\right\}^2
+\frac{\E^{\I\zeta s^{(\nu)}\sigma_z}}{2\left.s^{(\nu)}\right.^2}
\left[
\sin\left(2\zeta s^{(\nu)}\right)-2\zeta s^{(\nu)}
\right]\I\sigma_z
\nonumber\\
&=&
\frac{\E^{\I\zeta s^{(\nu)}\sigma_z}}{2\left.s^{(\nu)}\right.^2}
\left\{
1-\cos2\zeta s^{(\nu)}
+\left[\sin2\zeta s^{(\nu)}-2\zeta s^{(\nu)}
\right]\I\sigma_z
\right\},
\end{eqnarray}
and hence arrive at the expression
\begin{eqnarray}\label{eq3400}
\lefteqn{
T^{(\nu)}
=\E^{\I\zeta s^{(\nu)}\sigma_z}
\left\{
1+\frac{r^{(\nu)}}{2s^{(\nu)}}
\left\{
\sigma_x\sin2\zeta s^{(\nu)}
+\sigma_y\left[1-\cos2\zeta s^{(\nu)}\right]
\right\}
\phantom{\frac{\left.r^{(\nu)}\right.^2}{4\left.s^{(\nu)}\right.^2}}\right.
}
\nonumber\\
&&\left.
+\frac{\left.r^{(\nu)}\right.^2}{4\left.s^{(\nu)}\right.^2}
\left\{
1-\cos2\zeta s^{(\nu)}
+\left[\sin2\zeta s^{(\nu)}-2\zeta s^{(\nu)}
\right]\I\sigma_z
\right\}
\right\}
\nonumber\\
&=&
\left(\begin{array}{cc}
\E^{\I\zeta s^{(\nu)}} & 0\\
0 & \E^{-\I\zeta s^{(\nu)}}
\end{array}\right)\times
\nonumber\\
&&
\left(\begin{array}{cc}
1+\displaystyle{\frac{\left.r^{(\nu)}\right.^2}{4\left.s^{(\nu)}\right.^2}}
\left(
1-\E^{-2\I\zeta s^{(\nu)}}-2\I\zeta s^{(\nu)}
\right)
&
\displaystyle{\frac{-\I r^{(\nu)}}{2s^{(\nu)}}}
\left(
1-\E^{-2\I\zeta s^{(\nu)}}
\right)
\\
\displaystyle{\frac{\I r^{(\nu)}}{2s^{(\nu)}}}
\left(\begin{array}{l}
1-\E^{2\I\zeta s^{(\nu)}}
\end{array}\right)
&
1+\displaystyle{\frac{\left.r^{(\nu)}\right.^2}{4\left.s^{(\nu)}\right.^2}}
\left(
1-\E^{2\I\zeta s^{(\nu)}}+2\I\zeta s^{(\nu)}
\right)
\end{array}\right)
\nonumber\\
&=&\left(\begin{array}{cc}
\E^{\I\zeta s^{(\nu)}}+\displaystyle{\frac{\I\left.r^{(\nu)}\right.^2}{2\left.s^{(\nu)}\right.^2}}
\left[\sin\left(\zeta s^{(\nu)}\right)-\zeta s^{(\nu)}\E^{\I\zeta s^{(\nu)}}\right]
&
\displaystyle{\frac{r^{(\nu)}}{s^{(\nu)}}}
\sin\left(\zeta s^{(\nu)}\right)
\\
&
\\
\displaystyle{\frac{r^{(\nu)}}{s^{(\nu)}}}
\sin\left(\zeta s^{(\nu)}\right)
&
\E^{-\I\zeta s^{(\nu)}}-\displaystyle{\frac{\I\left.r^{(\nu)}\right.^2}{2\left.s^{(\nu)}\right.^2}}
\left[\sin\left(\zeta s^{(\nu)}\right)-\zeta s^{(\nu)}\E^{-\I\zeta s^{(\nu)}}\right]
\end{array}\right)
\qquad
\\ \label{eq3405}
&=&
\E^{\I\zeta s^{(\nu)}\sigma_z}
+\frac{r^{(\nu)}}{s^{(\nu)}}\sin\left(\zeta s^{(\nu)}\right)\sigma_x
+\frac{\I\left.r^{(\nu)}\right.^2}{2\left.s^{(\nu)}\right.^2}
\left[\sin\left(\zeta s^{(\nu)}\right)-\zeta s^{(\nu)}\E^{\I\zeta s^{(\nu)}\sigma_z}\right]\sigma_z,
\end{eqnarray}
which gives Eqs.~(\ref{eq3408}) and~(\ref{eq3409}).

Finally, we expand the transfer matrix with respect to $s^{(\nu)}/r^{(\nu)}$ in the limit opposite to the above.
We can actually use the expression~(\ref{eq3405}) of the expansion with respect to $r^{(\nu)}/s^{(\nu)}$.
Rotating the axes of the spin space around the spin $y$ axis, we have the transformation
\begin{equation}\label{eq3600}
\sigma_z\longrightarrow\sigma_x,
\quad
\sigma_x\longrightarrow-\sigma_z,
\qquad\mbox{and hence}\qquad
T^{(\nu)}=\E^{\I\zeta\left(s^{(\nu)}\sigma_z-\I r^{(\nu)}\sigma_x\right)}
\longrightarrow
\tilde{T}^{(\nu)}=\E^{\I\zeta\left(s^{(\nu)}\sigma_x+\I r^{(\nu)}\sigma_z\right)}.
\end{equation}
We can thereby obtain the expansion with respect to $s^{(\nu)}/r^{(\nu)}$ by replacing $s^{(\nu)}$ and $r^{(\nu)}$ in Eq.~(\ref{eq3405}) with $\I r^{(\nu)}$ and $\I s^{(\nu)}$, respectively.
The result is
\begin{equation}\label{eq3620}
\tilde{T}^{(\nu)}\simeq
\E^{-\zeta r^{(\nu)}\sigma_z}
+\frac{s^{(\nu)}}{r^{(\nu)}}\I\sinh\left(\zeta r^{(\nu)}\right)\sigma_x
+\frac{\I\left.s^{(\nu)}\right.^2}{2\left.r^{(\nu)}\right.^2}
\left[\I\sinh\left(\zeta r^{(\nu)}\right)-\I \zeta r^{(\nu)}\E^{-\zeta r^{(\nu)}\sigma_z}\right]\sigma_z.
\end{equation}
Then we rotate the axis back in the way opposite to Eq.~(\ref{eq3600}), arriving at
\begin{eqnarray}\label{eq3630}
T^{(\nu)}&\simeq&
\E^{\zeta r^{(\nu)}\sigma_x}
+\sigma_z\frac{\I s^{(\nu)}}{r^{(\nu)}}\sinh\zeta r^{(\nu)}
+\sigma_x\frac{\left.s^{(\nu)}\right.^2}{2\left.r^{(\nu)}\right.^2}
\left[\sinh\zeta r^{(\nu)}-\zeta r^{(\nu)}\E^{\zeta r^{(\nu)}\sigma_x}\right].
\nonumber\\
&=&
\left(\begin{array}{cc}
\begin{array}{l}
\displaystyle
\cosh\zeta r^{(\nu)}
+\frac{\I s^{(\nu)}}{r^{(\nu)}}\sinh\zeta r^{(\nu)}
\\
\displaystyle
-\frac{\left.s^{(\nu)}\right.^2}{2\left.r^{(\nu)}\right.^2}
\zeta r^{(\nu)}\sinh\zeta r^{(\nu)}
\end{array}
&
\begin{array}{l}
\displaystyle
\sinh\zeta r^{(\nu)}
+\frac{\left.s^{(\nu)}\right.^2}{2\left.r^{(\nu)}\right.^2}
\sinh\zeta r^{(\nu)}
\\
\displaystyle
-\frac{\left.s^{(\nu)}\right.^2}{2\left.r^{(\nu)}\right.^2}
\zeta r^{(\nu)}\cosh\zeta r^{(\nu)}
\end{array}
\\
&
\\
&
\\
\begin{array}{l}
\displaystyle
\sinh\zeta r^{(\nu)}
+\frac{\left.s^{(\nu)}\right.^2}{2\left.r^{(\nu)}\right.^2}
\sinh\zeta r^{(\nu)}
\\
\displaystyle
-\frac{\left.s^{(\nu)}\right.^2}{2\left.r^{(\nu)}\right.^2}
\zeta r^{(\nu)}\cosh\zeta r^{(\nu)}
\end{array}
&
\begin{array}{l}
\displaystyle
\cosh\zeta r^{(\nu)}
-\frac{\I s^{(\nu)}}{r^{(\nu)}}\sinh\zeta r^{(\nu)}
\\
\displaystyle
-\frac{\left.s^{(\nu)}\right.^2}{2\left.r^{(\nu)}\right.^2}
\zeta r^{(\nu)}\sinh\zeta r^{(\nu)}
\end{array}
\end{array}\right),
\end{eqnarray}
which gives Eqs.~(\ref{eq3648}) and~(\ref{eq3649}).

\end{document}